\documentclass[conference, 10pt]{IEEEtran}
\IEEEoverridecommandlockouts

\usepackage{url}
\usepackage{cite}
\usepackage{amsmath,amssymb,amsfonts}
\usepackage{algorithmic}
\usepackage{graphicx}
\usepackage{textcomp}
\usepackage{xcolor}
\def\BibTeX{{\rm B\kern-.05em{\sc i\kern-.025em b}\kern-.08em
    T\kern-.1667em\lower.7ex\hbox{E}\kern-.125emX}}

\usepackage{enumitem}

\usepackage{romannum}
\usepackage{textcomp}
\usepackage{tabularx}
\usepackage{diagbox}
\usepackage{multirow}
\usepackage{hhline}

\usepackage{bbding}
\usepackage{pifont}
\usepackage{wasysym}

\usepackage{cleveref}
\crefformat{section}{\S#2#1#3}
\crefformat{subsection}{\S#2#1#3}
\crefformat{subsubsection}{\S#2#1#3}

\usepackage{listings}
\usepackage{color}
\definecolor{dkgreen}{rgb}{0,0.6,0}
\definecolor{gray}{rgb}{0.5,0.5,0.5}
\definecolor{mauve}{rgb}{0.58,0,0.82}
\definecolor{backcolour}{rgb}{0.95,0.95,0.92}
\usepackage{graphics}
\usepackage{caption}
\usepackage{subcaption}
\graphicspath{{./figures/}}

\usepackage{amsmath}
\usepackage{mathrsfs}

\begin{document}

\title{FLiCR: A Fast and Lightweight LiDAR Point Cloud Compression Based on Lossy RI}
\author{}

\author{\IEEEauthorblockN{Jin Heo}
\IEEEauthorblockA{\textit{Georgia Institute of Technology} \\
Atlanta, Georgia, USA \\
jheo33@gatech.edu}
\and
\IEEEauthorblockN{Christopher Phillips}
\IEEEauthorblockA{\textit{Adeia} \\
Hartwell, Georgia, USA \\
chris.phillips@adeia.com}
\and
\IEEEauthorblockN{Ada Gavrilovska}
\IEEEauthorblockA{\textit{Georgia Institute of Technology} \\
Atlanta, Georgia, USA \\
ada@cc.gatech.edu}
}

%\renewcommand\footnotetextcopyrightpermission[1]{}

% Additional packages

\pagenumbering{arabic}

\newcommand{\jin}[1]{\textcolor{blue}{JH: #1}}
\newcommand{\ada}[1]{\textcolor{red}{AG: #1}}
\newcommand{\blue}[1]{\textcolor{blue}{#1}}
\newcommand{\red}[1]{\textcolor{red}{#1}}
\newcommand{\green}[1]{\textcolor{green}{#1}}
\newcommand{\mul}{$\times$}

\newenvironment{tightitemize}%
 {\begin{list}{$\bullet$}{%
 		\setlength{\leftmargin}{10pt}
        \setlength{\itemsep}{0pt}%
        \setlength{\parsep}{0pt}%
        \setlength{\topsep}{0pt}%
        \setlength{\parskip}{0pt}%
        }%
 }%
{\end{list}}

\newcommand{\specialcell}[2][c]{%
  \begin{tabular}[#1]{@{}c@{}}#2\end{tabular}}

\maketitle
\thispagestyle{plain}
\pagestyle{plain}

\begin{abstract}
Light detection and ranging (LiDAR) sensors are becoming available on modern mobile devices and provide a 3D sensing capability.
This new capability is beneficial for perceptions in various use cases, but it is challenging for resource-constrained mobile devices to use the perceptions in real-time because of their high computational complexity.
In this context, edge computing can be used to enable LiDAR online perceptions, but offloading the perceptions on the edge server requires a  low-latency, lightweight, and efficient compression due to the large volume of LiDAR point clouds data.

This paper presents FLiCR, a fast and lightweight LiDAR point cloud compression method for enabling edge-assisted online perceptions.
FLiCR is based on range images (RI) as an intermediate representation (IR), and dictionary coding for compressing RIs.
FLiCR achieves its benefits by leveraging lossy RIs, and we show the efficiency of bytestream compression is largely improved with quantization and subsampling.
In addition, we identify the limitation of current quality metrics for presenting the entropy of a point cloud, and introduce a new metric that reflects both point-wise and entropy-wise qualities for lossy IRs.
The evaluation results show FLiCR is more suitable for edge-assisted real-time perceptions than the existing LiDAR compressions, and we demonstrate the effectiveness of our compression and metric with the evaluations on 3D object detection and LiDAR SLAM.
\end{abstract}

{\let\thefootnote\relax\footnote{{© 2022 IEEE.  Personal use of this material is permitted.  Permission from IEEE must be obtained for all other uses, in any current or future media, including reprinting/republishing this material for advertising or promotional purposes, creating new collective works, for resale or redistribution to servers or lists, or reuse of any copyrighted component of this work in other works.}}}

\begin{IEEEkeywords}
lidar, lidar point cloud, lidar point cloud compression, 3D point cloud compression, remote lidar perceptions, real-time perception service, range image compression, edge computing
\end{IEEEkeywords}

%------------------------------------------------------------------------------
\section{Introduction}
\label{sec:intro}
%------------------------------------------------------------------------------

% General Context about LiDAR
Light detection and ranging (LiDAR) sensors have been used in robotics and autonomous vehicles for robust and accurate 3D sensing.
The 3D point clouds from LiDAR sensors are used in perception tasks for understanding real-world contexts, such as 3D object detection and tracking, and LiDAR simultaneous localization and mapping (LiDAR SLAM).
A LiDAR sensor has advantages over image sensors as it can provide environmental information in 3D with higher accuracy and is more robust to challenging weather and light conditions.
Previously, despite the benefits of LiDAR sensors, only a few device platforms were equipped with them because the early models of LiDAR were too expensive and large in size~\cite{expensive}.
However, as LiDAR technology advances, the sensors are becoming smaller and affordable, while maintaining their sensing performance even at lower power usage.
Recently, Velodyne Lidar released a palm-size LiDAR sensor, Velabit~\cite{velabit}, for around \$100, Intel RealSense camera~\cite{realsense} has a tiny LiDAR sensor, as do the latest Apple iPhone and iPad~\cite{applelidar} devices.
As LiDAR becomes cost-effective and smaller in size, there would be more opportunities for mobile devices to leverage perceptions of this new sense in diverse use cases.

% General Problem
Although LiDAR technology is becoming available on mobile devices, there remain challenges in leveraging LiDAR perceptions for real-time use cases.
As discussed in prior LiDAR perception research~\cite{simon2019complexer, simony2018complex, ye2020hvnet, yang2018pixor, zhang2020polarnet}, the computational complexities of LiDAR perceptions are very high because these algorithms process unstructured 3D data.
Even with these highly optimized algorithms, high-end processors and GPUs are required to make them run in real-time.
This poses limitations to enabling real-time LiDAR perception on resource-constrained (including by battery lifetime) mobile devices which lack such high-end hardware.

% General Solution
Edge computing is a technology that can relieve such issues and enable computationally intensive perceptions for mobile users with commodity hardware.
The edge (or cloudlet) is located at the edge of the network and close to the end users in the multi-tier cloud.
End users can utilize edge resources effectively for storing and processing data on nearby edge servers accessible via low-latency and high-bandwidth networks such as 5G~\cite{khan2019edge}.
There has been research on offloading image-based real-time perceptions on the edge~\cite{chen2015glimpse, liu2019edge, jin2021flexible}.
When remotely running the perceptions, a sequence of images is transmitted to the edge server in real-time.
In such settings, efficient and low-latency image compression is essential because of the large size of the raw images.
Fortunately, image compression algorithms have been extensively studied by both industry and academia~\cite{richardson2004h, grange2016vp9, sullivan2012overview}, and the accelerators for such standard codecs are broadly available even on mobile devices, given the popularity of video streaming~\cite{intelquicksink, snapd, nvcodec}.
These existing video codecs with accelerators enable the online perceptions on the remote edge servers.

% Specific Problem: LiDAR Compression
While offloading image-based perceptions takes advantage of the available codecs and accelerators for real-time video streaming, there are challenges to enabling  edge-assisted online LiDAR perceptions due to the lack of such infrastructure for LiDAR point cloud compression.
LiDAR sensors generate unstructured 3D point clouds, and their volume is too large to send them as raw data.
As an example, the points per scan of the KITTI dataset~\cite{geiger2013vision} are about 120,000 of 2 MB, and streaming the raw sensor data at 60 frames per second (FPS) needs a bandwidth of 120 Mbps.
In addition to the high bandwidth usage, transmitting a large amount of data not only causes higher network loads on the backend middleboxes, resulting in additional transmission delays, but also reduces the lifetime of a mobile device by consuming its battery~\cite{xiao2013modeling, vergara2013energybox, zhang2012mili}.
In this context, an effective point cloud compression is crucial, but it is required to be low-latency and lightweight. Since the responsiveness of online perceptions is determined by the end-to-end latency, high compression latency can introduce  discrepancy between the real environment and the perception results~\cite{li2020towards}.
The compression time should be sufficiently small not to compromise the benefit of the reduced perception processing time on the server.
Moreover, the compression should be lightweight to run on mobile devices. % compresses the point clouds.

There are prior efforts to improve 3D point cloud compression, but
their primary focus is  on achieving higher compression ratio
while preserving the original content qualities~\cite{mammou2019g, tu2016compressing, tu2019point, ahn2014large, houshiar20153d, huang2020octsqueeze, graziosi2020overview}.
Even with real-time compression methods, the latency ranges of previous methods are too high to enable online perceptions,
or they require  high-end processors with GPUs for low latency~\cite{feng2020real, tu2019real, que2021voxelcontext, sun2020novel, sun2019novel, song2021layer}.
In short, the effectiveness of existing point cloud compression methods is limited when considering mobile devices, which poses a challenge to enabling edge-assisted LiDAR perceptions.

For enabling edge-assisted LiDAR perceptions to real-time applications, we propose \textbf{FLiCR}, a lightweight and low-latency point cloud compression method based on the range-image (RI) representation and a lossless compression algorithm.
While previous research on RI compressions utilizes only the quantization of the point bit precision with lossless RI mapping~\cite{ahn2014large, feng2020real, houshiar20153d, tu2016compressing, tu2019point, tu2019real}, we explore the optimization opportunities of lossy RIs with subsampling for reducing the data size and improving the compression efficiency.
The idea is that compression algorithms such as dictionary coding, which use shorter references to repetitive features, would become more effective since they operate with a more limited data representation space.
Subsampling of mapped points leads to point loss, and it is criticial to understand how this translates to reduction in end-to-end perception quality.
We demonstrate the limitations of existing quality metrics to represent this total information loss because their designs are only concerned with point-to-point distances.
Then, we propose a unified metric, \textbf{ePSNR}, that captures both point-wise and entropy-wise point cloud qualities, by extending the current PSNR with a probability function of entropy estimation.

We evaluate FLiCR and ePSNR with the current compression methods and metrics and different LiDAR perception use cases.
In our results, FLiCR achieves up to 5.3$\times$ improvement in end-to-end compression latency on mobile devices and 12.6$\times$ in compression ratio compared to Google Draco, and ePSNR captures the quality impact of the lossyness introduced by FLiCR, enabling future system support to dynamically exercise the latency-performance tradeoff it exposes.
In summary, we make the following contributions:
\begin{tightitemize}
  \item We identify the requirements of LiDAR point cloud compression methods for edge-assisted online perceptions and conduct a thorough analysis on the limitations of the state-of-the-art technologies.
  \item We propose FLiCR, a lightweight, low-latency, and efficient compression method that combines use of lossy RI and lossless dictionary coder, and compare it to the existing methods.
  \item We point out the limitations of the current quality metrics for point clouds in terms of the entropy loss and propose ePSNR as a new single-number metric reflecting both point-wise and entropy-wise qualities.
  \item We demonstrate the benefits of our compression method and metric on two downstream perception tasks, 3D object detection and LiDAR SLAM.
\end{tightitemize}

%------------------------------------------------------------------------------
\section{Background}
\label{sec:background}
%------------------------------------------------------------------------------

\noindent\textbf{3D Point Clouds.\quad} A 3D point cloud is a set of points in the 3D space.
Point clouds can be categorized into two categories by their characteristics: structured and unstructured.
The unstructured (raw) point cloud is a sequence of the coordinate values of 3D points (usually \emph{x, y, z} in a Cartesian coordinate system), optionally with other attributes such as reflection intensities.
The structured point cloud is a point set organized with geometric or hierarchical structure contexts including meshes, octrees, etc.
A LiDAR point cloud is an unstructured point cloud directly captured from LiDAR sensors.

\begin{figure}[b]
  \centering
  \includegraphics[width=0.8\linewidth]{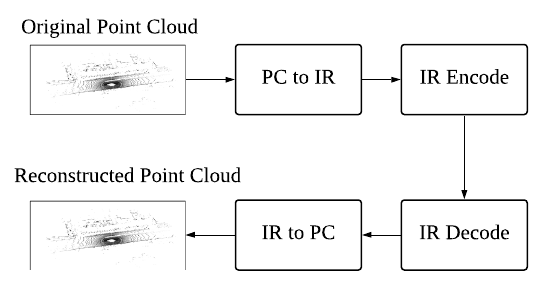}
  \caption{\small The general processing steps of the existing point cloud compression methods.}
  \label{fig:pcprocess}
\end{figure}

\begin{figure}
  \centering
  \begin{subfigure}[t]{0.4\textwidth}
    \includegraphics[width=\textwidth]{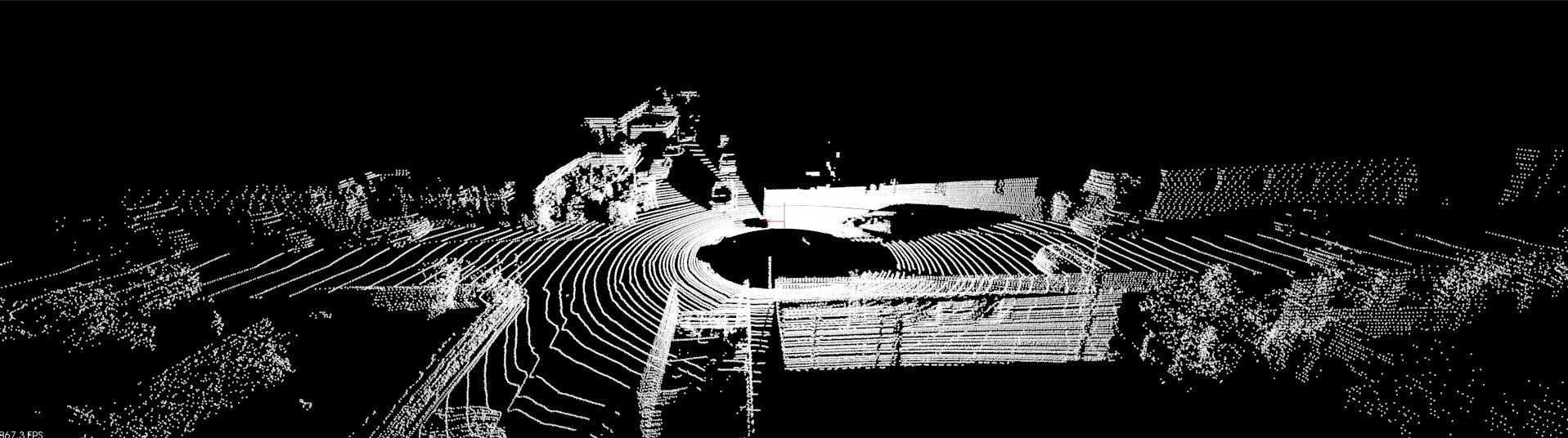}
    \caption{\small A raw point cloud.}
    \label{fig:rawpcrep}
  \end{subfigure}\hspace{0.02\textwidth}
  \begin{subfigure}[t]{0.4\textwidth}
    \includegraphics[width=\textwidth]{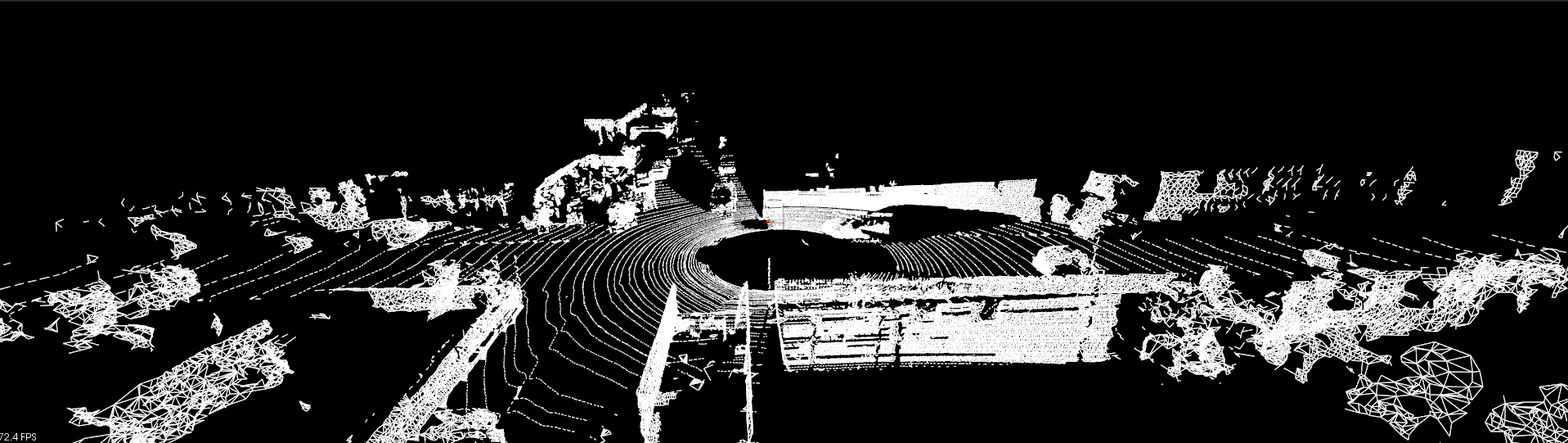}
    \caption{\small A mesh representation.}
    \label{fig:meshrep}
  \end{subfigure}

  \begin{subfigure}[t]{0.4\textwidth}
    \includegraphics[width=\textwidth]{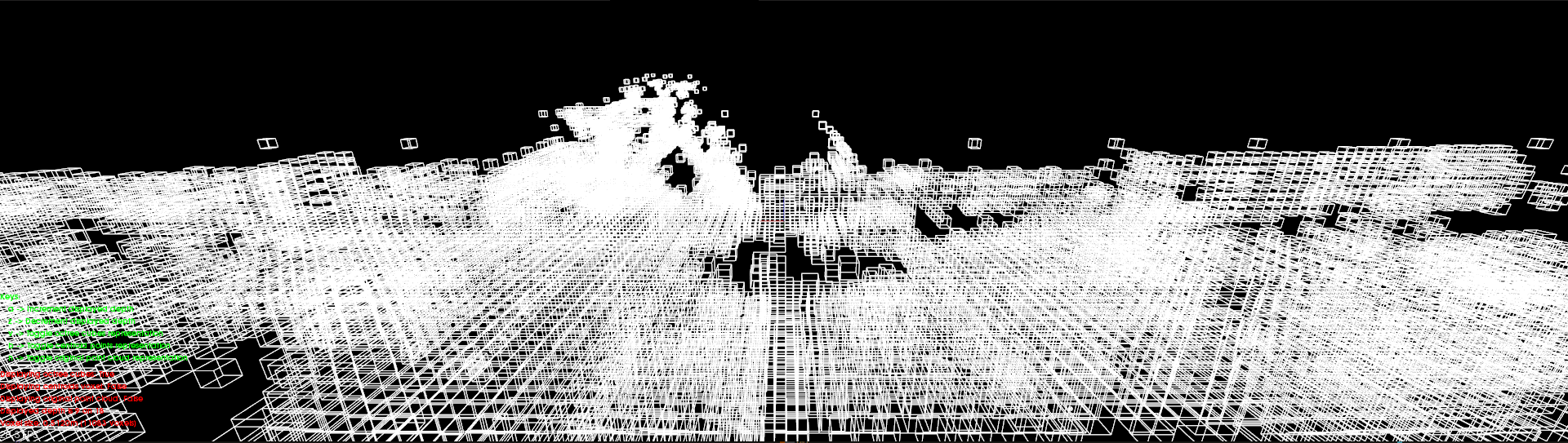}
    \caption{\small An octree representation.}
    \label{fig:octreerep}
  \end{subfigure}\hspace{0.02\textwidth}
  \begin{subfigure}[t]{0.4\textwidth}
    \includegraphics[width=\textwidth]{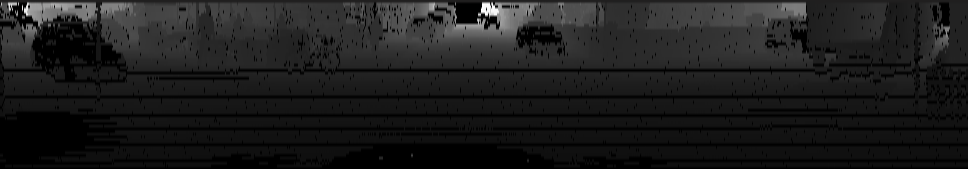}
    \caption{\small A range image.}
    \label{fig:rirep}
  \end{subfigure}
  \caption {\small The visualizations of a LiDAR point cloud from KITTI~\cite{geiger2013vision} dataset with different IRs.}
  \label{fig:reps}
\end{figure}

\noindent\textbf{Unstructured Point Cloud Compression.\quad} There are
diverse existing compression methods, but a common thread across them
is to convert raw point clouds into structured intermediate
representations (IRs) and apply compression algorithms to the IRs, as
shown in Figure~\ref{fig:pcprocess}.
The compression process is tied to each IR, and the commonly used IRs are k-d tree, octree, mesh, and range image.
Figure~\ref{fig:reps} shows different IR visualizations from a raw point cloud.
Compression methods are categorized into geometry-based or image-based
compression, based on the used IRs.
Geometry-based compression uses the tree structures or mesh~\cite{rusu20113d, devillers2000geometric, mammou2019g, que2021voxelcontext, huang2020octsqueeze, biswas2020muscle, nguyen2021multiscale, draco}, and the image-based compression maps the point clouds into 2D frames~\cite{feng2020real, tu2016compressing, tu2019point, tu2019real, sun2019novel}.
The geometry-based compressions code their IRs and compress the coded IRs, and the image-based approaches utilize the existing codecs or present their own techniques for compressing the mapped images.
More details of the existing methods appear in Section~\ref{sec:related}.

%------------------------------------------------------------------------------
\section{Challenges with L\lowercase{i}DAR Point Cloud Compression}
\label{sec:motivation}
%------------------------------------------------------------------------------

\begin{figure}
  \centering
  \begin{subfigure}[t]{0.23\textwidth}
    \includegraphics[width=\textwidth]{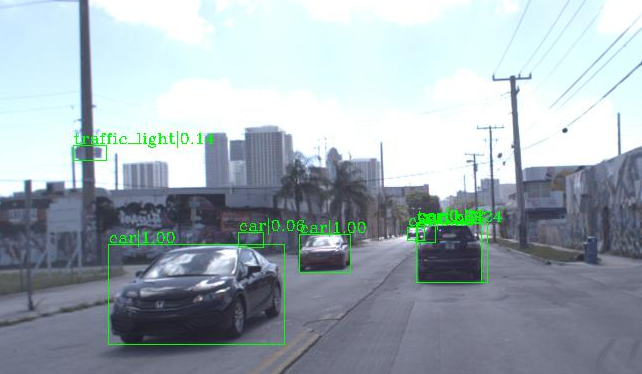}
    \caption{\small The object detection result without the discrepancy latency.}
    \label{fig:odgt}
  \end{subfigure}\hspace{0.02\textwidth}
  \begin{subfigure}[t]{0.23\textwidth}
    \includegraphics[width=\textwidth]{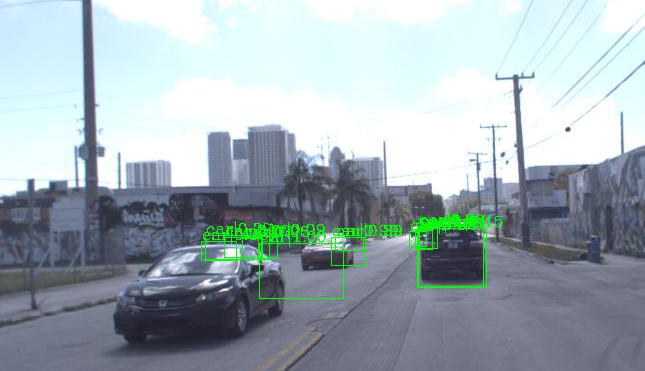}
    \caption{\small The object detection result with the discrepancy latency of 300 ms.}
    \label{fig:od300ms}
  \end{subfigure}
  \caption {\small The simulated results of the online perceptions with Mask R-CNN~\cite{he2017mask} and Argoverse~\cite{chang2019argoverse}.}
  \label{fig:odres}
\end{figure}

\noindent{\bf Reducing discrepancy latency. }
For online perception, the end-to-end latency of the processing pipeline has a major impact on performance as it affects the application responsiveness to changes in the real-world environment~\cite{li2020towards}.
For instance, if the perception result of the data captured at $t_{0}$ is available at $t_{1}$, there would be a discrepancy between the result and real world with the changes during the time from $t_{0}$ to $t_{1}$.
Figure~\ref{fig:odres} is the screenshot of the simulated online perception results with and without 300 ms of the discrepancy latency in the object detection task.
Without the discrepancy latency, all real-world objects are aligned with the detection results as Figure~\ref{fig:odgt}.
However, with the latency, the perception results are not correctly corresponding to the real-world objects because of the discrepancies as Figure~\ref{fig:od300ms}.

\begin{figure}[]
  \centering
  \includegraphics[width=0.8\linewidth]{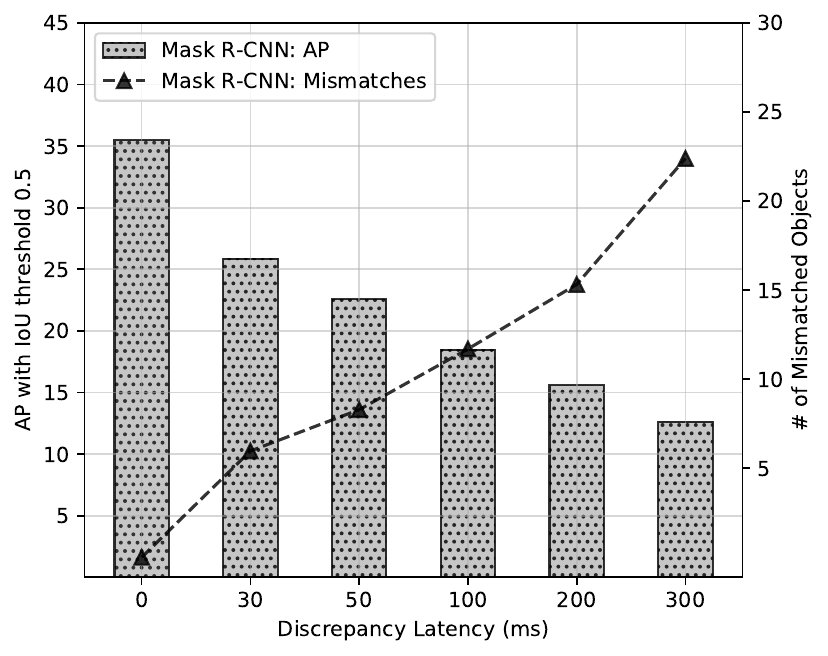}
  \caption{\small The measured AP and number of mismatches of the online object detection with different discrepancy latencies.}
  \label{fig:sap}
\end{figure}

In Figure~\ref{fig:sap}, we show the impact of different discrepancy latencies between the real world and the perception results on the performance of object detection.
We use the metrics for streaming perceptions~\cite{li2020towards} -- the average precision (AP) with intersection over union (IoU) threshold 0.5, and the number of mismatched objects between the results with and without discrepancy latencies.
For object detection, we use Mask R-CNN~\cite{he2017mask} pre-trained with the dataset of Microsoft COCO~\cite{lin2014microsoft}.
Then, the detection model runs with an autonomous driving dataset, Argoverse~\cite{chang2019argoverse}, and we measure the metrics with different discrepancy latencies.
Without the discrepancy latency, the perception model detects $\sim$36\% of objects to the ground truth.
As the latency increases, the AP result starts to decrease with the increased number of mismatched objects.
These results show the latency governs the perception performance and there is the latency-performance tradeoff of online perceptions.

In addition to the perception algorithm, there are additional components in the processing pipeline which introduce latency when offloading the LiDAR perceptions on the edge: data compression and network transportation.
The overheads from these steps contribute to the discrepancy latency, and the benefit of the reduced processing time of a perception can be compromised by them.
As shown in Figure~\ref{fig:sap}, even with 30 ms of discrepancy latency, AP decreases by $\sim$28\% of the result without the latency.
With 100 ms delay, it becomes half of the result without delay.
Furthermore, the number of mismatched objects soars with the higher latencies.
So, along with the application of edge computing to reduce the network latency, a lightweight and low-latency LiDAR point cloud compression method is essential to enable edge-assisted online LiDAR perceptions for resource-constrained mobile users.

\begin{table*}[]
  \caption{\label{tab:expcc} The benchmark results of the existing compression methods for 3D point clouds. The results in parentheses are on the Jetson AGX.}
  \begin{tabularx}{\textwidth}{{|>{\raggedright}X
                                |>{\centering}c
                                |>{\centering}c
                                |>{\centering\arraybackslash}c
                                |>{\centering\arraybackslash}c
                                |>{\centering\arraybackslash}c
                                |>{\centering\arraybackslash}c
                                |>{\centering\arraybackslash}c
                                |>{\centering\arraybackslash}c|}}
  \hhline{=======}
    \diagbox[width=18.6em]{Metrics}{Methods}      & RLE          & Dict Coding~\cite{ziv1977universal}  & Google Draco~\cite{draco}    & MPEG G-PCC~\cite{mammou2019g}  & PCL~\cite{rusu20113d}    & RT-ST~\cite{feng2020real}  \\ \hline
    Compression Ratio                             & 0.54         & 1.67                                 & \textbf{17.05}               & 8.76                           & 5.72                     & 15.96                      \\ \hline
    PSNR (dB)                                     &\textbf{Lossless}      & \textbf{Lossless}           & 67.29                        & 78.43                          & 89.77                    & 63.18                      \\ \hline
    CD (cm)                                       &\textbf{Lossless}      & \textbf{Lossless}           & 0.267                        & 0.184                          & 0.001                    & 3.07                       \\ \hline
    Enc Time (ms)                                 & 40.1 (42.4)           & 40.5 (75.4)                 & \textbf{21.1} (\textbf{48.4})& 598 (741)                      & 72.1 (198)               & 97.7 (240)                 \\ \hline
    Dec Time (ms)                                 & 17.9 (15.8)           & 13.8 (33.7)                 & \textbf{9.44} (\textbf{18.6})& 204 (265)                      & 55.4 (153)               & 15.2 (34.8)                \\ \hline
    Enc Energy Usage (J)                          & 1.18 (\textbf{0.11})  & 1.19 (0.23)                 & \textbf{0.83} (0.14)         & 15.35 (2.41)                   & 2.05 (0.46)              & 2.63 (0.59)                \\ \hline
    Dec Energy Usage (J)                          & 0.51 (\textbf{0.04})  & 0.39 (0.08)                 & \textbf{0.36} (0.05)         & 5.56 (0.66)                    & 1.54 (0.36)              & 0.57 (0.12)                \\
  \hhline{=======}
  \end{tabularx}
\end{table*}

\noindent{\bf Limitations of existing compression methods. }
Given the popularity of 3D point cloud data, there are existing technologies for LiDAR point cloud compression: Google Draco~\cite{draco}, MPEG Geometry based point cloud compression (G-PCC)~\cite{mammou2019g}, Point Cloud Library (PCL)~\cite{rusu20113d} octree compression, and the real-time spatio-temporal (RT-ST) compression by Feng \emph{et al.}~\cite{feng2020real}.
Google Draco is based on k-d tree, PCL and G-PCC are on octree, and
 RT-ST compresses range images (RIs) of 3D point clouds.
Moreover, with these methods, it is possible to apply bytestream compressions of run-length encoding (RLE) and dictionary coding (LZ77)~\cite{ziv1977universal} to the point clouds directly; they treat point clouds as raw byte arrays and deflate them losslessly.
We compare these methods based on performance, quality, and efficiency metrics on  a desktop and on NVIDIA Jetson AGX of our experimental testbed, as described in Section~\ref{sec:subsec_exp}.
As the quality metrics of the point clouds, we use peak signal-to-noise ratio (PSNR) and Chamfer Distance (CD) as defined in Section~\ref{sec:metric}.
For each method, we encode and decode 100 point clouds from the KITTI dataset~\cite{geiger2013vision}.
The averaged results are in Table~\ref{tab:expcc}, showing the results of the desktop, and in parentheses the results from the Jetson.

While every method has its advantages for different metrics, we focus on the compression ratio, energy usage and latency, because these metrics show how well a compression method meets the requirements of point cloud compression for edge-assisted online perceptions.
For our target use case, there are three requirements of a compression method.

\begin{enumerate}[noitemsep,topsep=0pt,parsep=0pt,partopsep=0pt,label=(\arabic*),leftmargin=16pt]
\item It should be very low-latency because of the latency-performance tradeoff of online perceptions;
\item The compression performance effectiveness in reducing the data size is important since the larger size of compressed data causes higher network cost and energy consumption on the client to transmit;
\item It should be lightweight to run on mobile devices of limited resources while satisfying the other requirements.
\end{enumerate}

For the compression ratio and latency, Google Draco outperforms the other methods and is highly efficient in terms of the compression ratio and energy usage.
Although RLE shows comparable latency and energy usage on Jetson, it shows  increased total size when applying RLE directly to the floating-point values of 3D points.
Except for the lossless methods, PCL's octree compression shows the highest quality metrics.
However, it is at the cost of the lower efficiency and high energy usage versus Google Draco.

Based on the results in Table~\ref{tab:expcc}, Google Draco seems the best option to meet the requirements, but it causes about 60 ms ($\sim$50 ms for encoding and $\sim$10 ms for decoding) of the compression cost when the user device is Jetson and our desktop is a server.
Since there are additional delays from network transmission and algorithm processing, the discrepancy latency of the whole pipeline will be too high given a compression cost of 60 ms.
This would hurt the perception performance and be hard to use,
as illustrated in Figure~\ref{fig:sap}.

In summary, {\em there is a need for a low-latency, lightweight, and efficient LiDAR compression method for edge-assisted online perceptions}, which motivates us to pursue this work.

%------------------------------------------------------------------------------
\section{Intermediate Representations for FL\lowercase{i}CR}
\label{sec:ir}
%------------------------------------------------------------------------------

For meeting the aforesaid requirements, it is important to select a proper IR because the compression is dependent on each IR.
In this section, we microbenchmark the IR conversions and point out the benefit of range images (RIs) over the others in the context of enabling remote online perceptions.

\begin{table}[htbp]
  \caption{The latencies (ms) of each IR construction.}
  \begin{center}
  \begin{tabular}{ |c|c|c|c|c|c| }
    \hline
             & RI    & Parallel RI   & Octree & K-d tree & Mesh \\ \hline
     Desktop & 11.78 & \textbf{6.72} & 30.67  & 13.21    & 1872 \\ \hline
     Jetson  & 16.34 & \textbf{9.26} & 32.11  & 32.44    & 2755 \\ \hline
  \end{tabular}
  \label{tab:riconv}
  \end{center}
\end{table}

Table~\ref{tab:riconv} shows the conversion latencies of the IRs with the LiDAR point clouds from the KITTI dataset~\cite{geiger2013vision}.
We use PCL~\cite{rusu20113d} implementations for octree and k-d tree,
and mesh conversion is based on the algorithm of Marton \emph{et al.}~\cite{marton2009fast}.
The RI conversion is our implementation, and the parallelized version is with OpenMP~\cite{dagum1998openmp}.
The RIs are generated by converting the raw points in the 3D Cartesian coordinates to the spherical coordinates.
Equation~\ref{eq:1} shows the conversion and $r$, $\theta$, and $\phi$ are the radial distance, polar angle, and azimuthal angle each.
When $\theta$ and $\phi$ are calculated, they are mapped to the frame pixel by the sensor's angular precisions.
For example, Velodyne HDL-64E used in the KITTI dataset has 0.08\textdegree~and 0.35\textdegree~for horizontal and azimuthal precisions with 360\textdegree~of the horizontal field of view (FoV) and 64 vertical lasers~\cite{hdl64}.
Thus, each scan's point cloud would be mapped to a 2D frame of 4500$\times$64.
By adjusting the parameters of precisions and FoV, RI can work on diverse LiDAR sensors.

\begin{equation}\label{eq:1}
\begin{aligned}
  r = \sqrt{x^2+y^2+z^2} \\
  \theta = arccos\left(\frac{z}{r}\right) \\
  \phi = arctan\left(\frac{y}{x}\right)
\end{aligned}
\end{equation}

In our results, the parallelized RI conversion shows the lowest latency on both desktop and Jetson as RI has advantages over other IRs in terms of simplicity and parallelism.
For the octree and k-d tree, there have been many efforts for their parallelized constructions~\cite{shevtsov2007highly, wehr2018parallel, lauterbach2009fast, karras2012maximizing, wu2011sah}.
However, as pointed out in the previous works, their hierarchical structures inherently make the construction processes sequential, and it is challenging to fully parallelize their constructions.
In contrast, the RI conversion can be easily parallelized since each point conversion of RI is completely independent from the others.
For the mesh, its generation from point clouds requires triangulation algorithms and calculating the surface normal for each mesh.
These processes require iterating each point and finding nearest
neighbors to generate a mesh, and the mesh conversion has high
computational complexity and is not suitable for real-time due to its large magnitude of execution time~\cite{marton2009fast, salman2010feature, guan2020voxel}.

Furthermore, these IRs have different theoretical complexities for the conversion.
The time complexities of the IR constructions are $O(n)$ for the RI and $O(n\log{}n)$ for the trees and mesh; the trees require  binary search for each point insertion, and the mesh construction needs nearest neighbor searches for the normal estimation and triangulation.
In addition, there is a side benefit that various image-processing techniques can be used for RIs.
Based on these observations, we adapt RI as the
target IR.

%------------------------------------------------------------------------------
\section{FL\lowercase{i}CR: Range Image Compression}
\label{sec:ricomp}
%------------------------------------------------------------------------------
Following selecting RI as the appropriate IR, the compression also needs to be efficient, low-latency, and lightweight.
In this section, we describe how to achieve the objectives of the compression method.
First, we identify the distortion issue of the current image codecs to LiDAR RIs.
Second, we explore the opportunity of lossy RIs for the downstream compression steps via RI quantization and subsampling.
We argue the lossless bytestream compressions can be hugely enhanced in terms of the compression efficiency and low latency through the lossy representation.
However, it compromises the point cloud quality, and we present the possible issues of FLiCR with lossy RIs.

\begin{table}[htbp]
  \caption{The compression ratios and qualities of H.264 with different QPs for 100 RIs of 4500$\times$64 and 8 bpp.}
  \begin{center}
  \begin{tabular}{ |c|c|c|c|c| }
    \hline
                            & QP 0    & QP 10 & QP 20  & QP 30 \\ \hline
    Compression Ratio       & 12.85   & 13.33 & 16.41  & 35.01 \\ \hline
    PSNR (dB)               & 63.18   & 48.21 & 37.61  & 38.12 \\ \hline
    CD (cm)                 & 2.23    & 16.25 & 126.06 & 107.16\\ \hline
  \end{tabular}
  \end{center}
  \label{tab:qpt}
\end{table}

\begin{figure}[htbp]
  \centering
  \begin{subfigure}[t]{0.38\textwidth}
    \includegraphics[width=\textwidth]{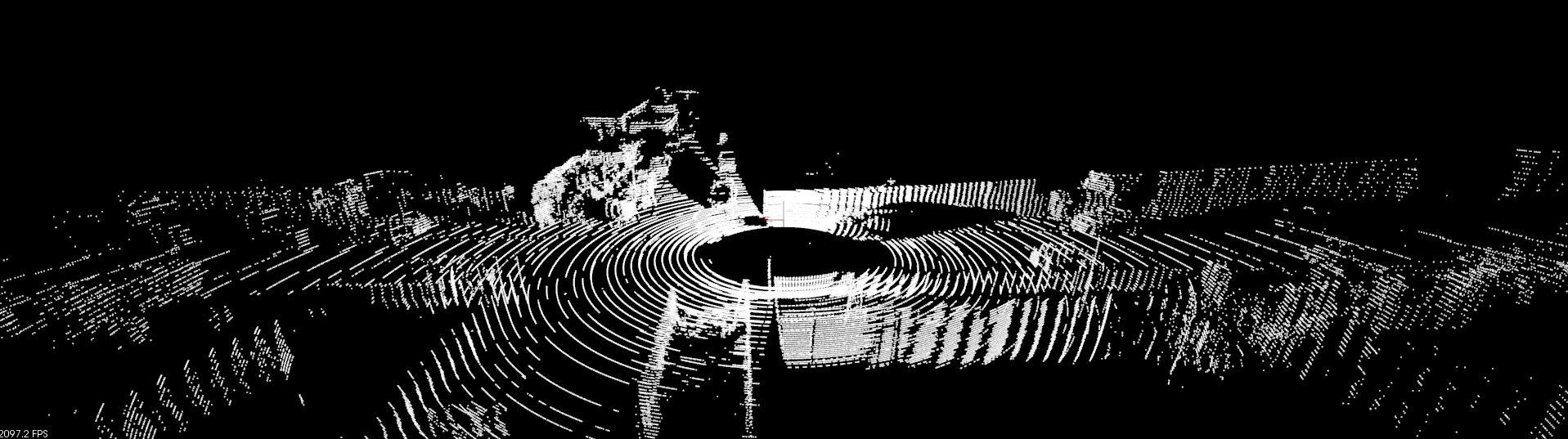}
    \caption{\small Point cloud of QP 0.}
    \label{fig:qp0}
  \end{subfigure}\hspace{0.02\textwidth}
  \begin{subfigure}[t]{0.38\textwidth}
    \includegraphics[width=\textwidth]{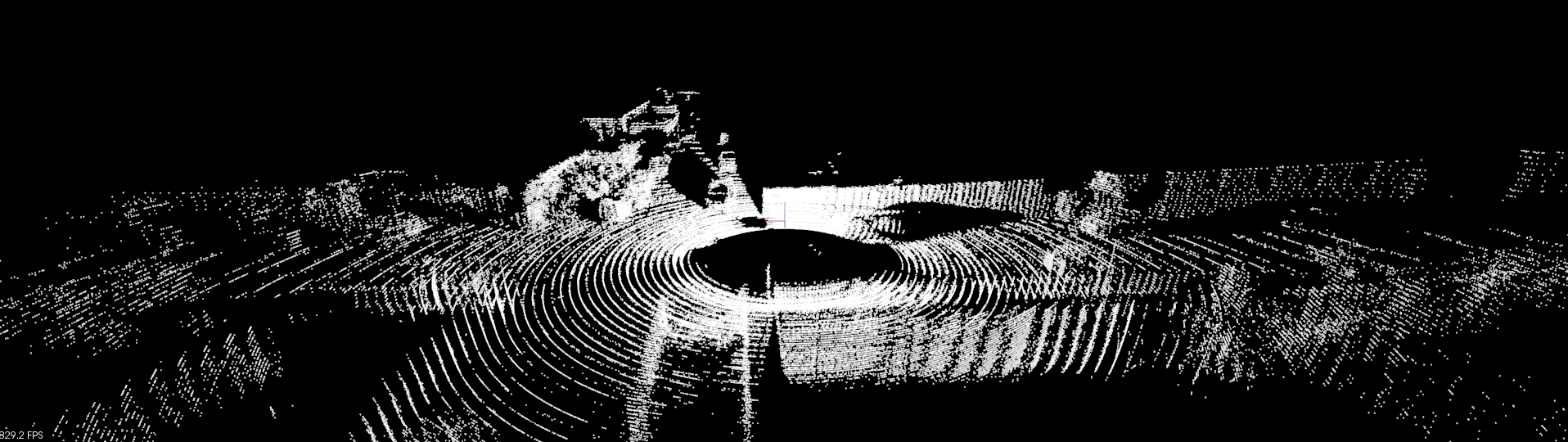}
    \caption{\small Point cloud of QP 10.}
    \label{fig:qp10}
  \end{subfigure}

  \begin{subfigure}[t]{0.38\textwidth}
    \includegraphics[width=\textwidth]{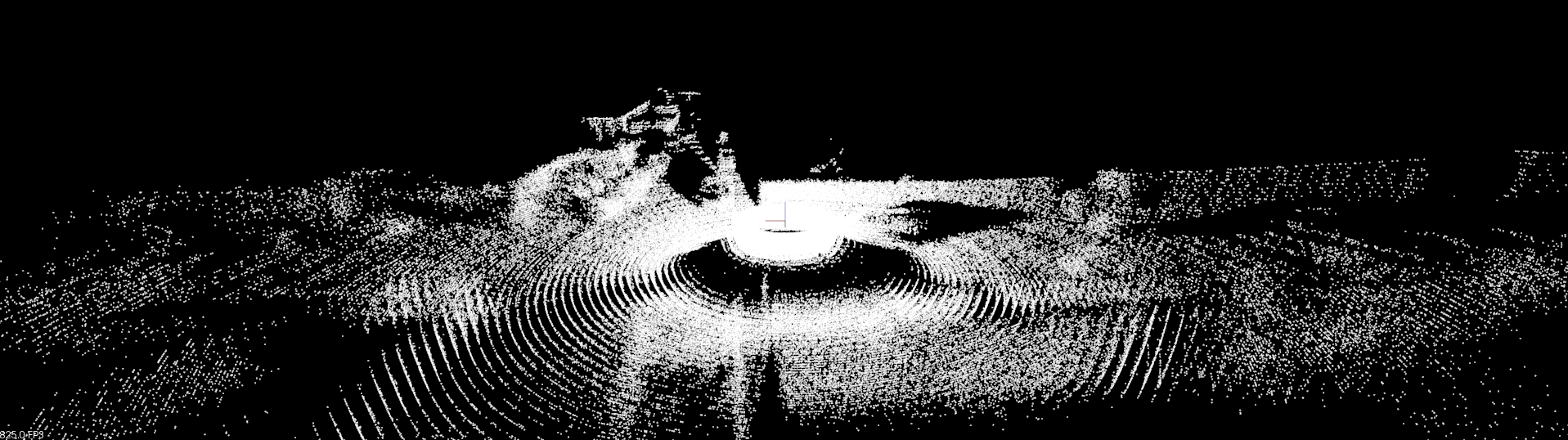}
    \caption{\small Point cloud of QP 20.}
    \label{fig:qp20}
  \end{subfigure}\hspace{0.02\textwidth}
  \begin{subfigure}[t]{0.38\textwidth}
    \includegraphics[width=\textwidth]{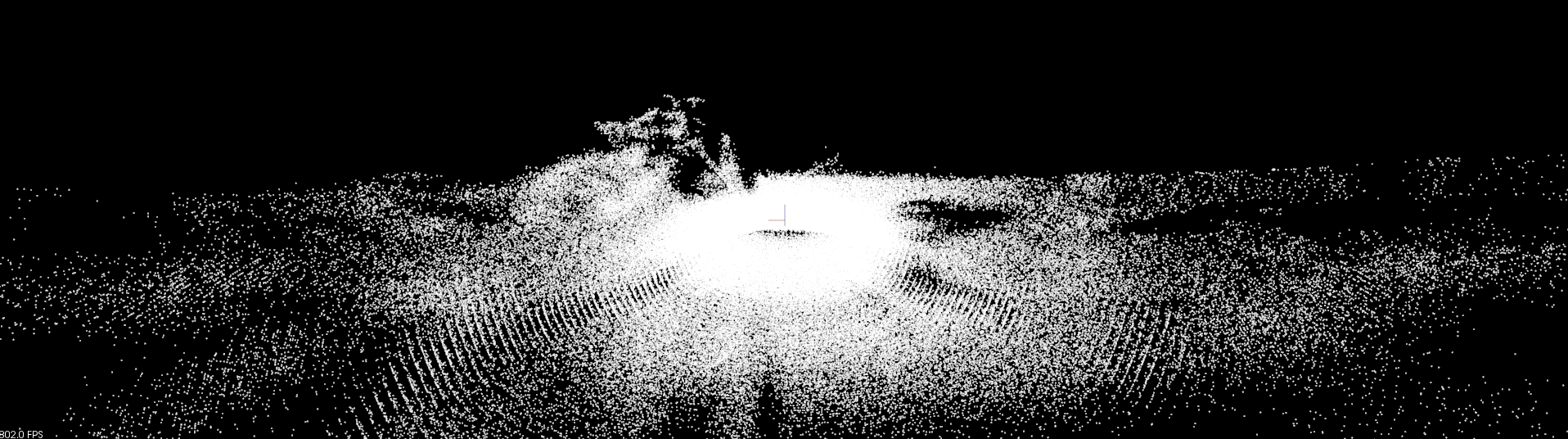}
    \caption{\small Point cloud of QP 30.}
    \label{fig:qp30}
  \end{subfigure}
  \caption {\small The visualizations of the reconstructed point clouds from the RI of 4500$\times$64 with H.264 and four different quantization parameters.}
  \label{fig:qp}
\end{figure}

\subsection{Issues with Current Image Compressions}
\label{sec:issuecodecs}
By representing LiDAR point clouds as images, it becomes possible to leverage the existing image-processing infrastructures and techniques.
With the popularity of video streaming, modern processor platforms and GPUs are equipped with dedicated hardware modules for the standard codecs such as H.264 and HEVC~\cite{intelquicksink, snapd, nvcodec}.
These codecs efficiently encode and decode continuous images with spatial and temporal optimizations~\cite{richardson2004h, sullivan2012overview}, and the pervasive accelerators enable the codecs in a low-latency and efficient way even with commodity mobile devices.

In this context, it seems appropriate to rely on the existing codecs with hardware accelerators at first glance.
However, we argue the existing codecs specialized for human vision are hardly applicable to RI compression.
The lossy image compression algorithms for human vision fully utilize
the characteristics of the human eyes to remove the data with minimal impacts to visual quality as much as possible; one example is to convert an image into the frequency domain via  discrete cosine transform (DCT) or fast Fourier transform (FFT) and the high frequency has more loss than the low-frequency data~\cite{richardson2004h, sullivan2012overview, marcellin2000overview}.
While the techniques that leverage the nature of human vision work well for normal images, the point cloud details are effectively lost as a result of the frequency-domain loss in LiDAR RIs.

Figure~\ref{fig:qp} shows the reconstructed point clouds from the RIs encoded and decoded via H.264 with different quantization parameters (QP).
QP regulates how much spatial detail is retained and is set from 0 for lossless to 51 for the most lossy compression.
As QP increases, the spatial detail is aggregated so that the encoded bit rate drops at the expense of data loss, resulting in lower quality~\cite{richardson2004h}.
The reconstructed point clouds become vague and noisy with high QPs.
Table~\ref{tab:qpt} shows the averaged compression ratio and quality metrics of the reconstructed point cloud with different QPs.
With the visual results of the reconstructed point clouds, the PSNR and CD results become worse drastically while the compression ratio increases moderately.
Considering that the quantization parameter such as QP or CRF of video streaming is usually set around 20 and 30 as a rule of thumb (FFmpeg's H.264 default CRF is 23~\cite{ffmpegh264}), these results show the current human-vision codecs are unsuitable for the RI compression.
For preserving the quality of point clouds, the codec quantization parameter should be set for lossless (QP 0), but it is at the cost of the lower compression efficiency, as Google Draco achieves $\sim$33\% higher compression ratio in Table~\ref{tab:expcc}.

\begin{table}[htbp]
  \caption{The existing quality metrics with sampling error (SE) for the subsampled RIs of 8 bpp.}
  \begin{center}
  \begin{tabular}{ |c|c|c|c|c| }
    \hline
               & 2048$\times$64    & 1024$\times$64 & 512$\times$64  & 256$\times$64 \\ \hline
     PSNR (dB) & 62.4              & 61.41          & 58.61          & 53.71         \\ \hline
     CD (cm)   & 5.37              & 9.23           & 15.22          & 36.17         \\ \hline
     SE        & 21.03\%           & 58.77\%        & 78.95\%        & 89.22\%       \\ \hline
  \end{tabular}
  \label{tab:oldmetrics}
  \end{center}
\end{table}

\begin{figure}[htbp]
  \centering
  \begin{subfigure}{0.38\textwidth}
    \includegraphics[width=\textwidth]{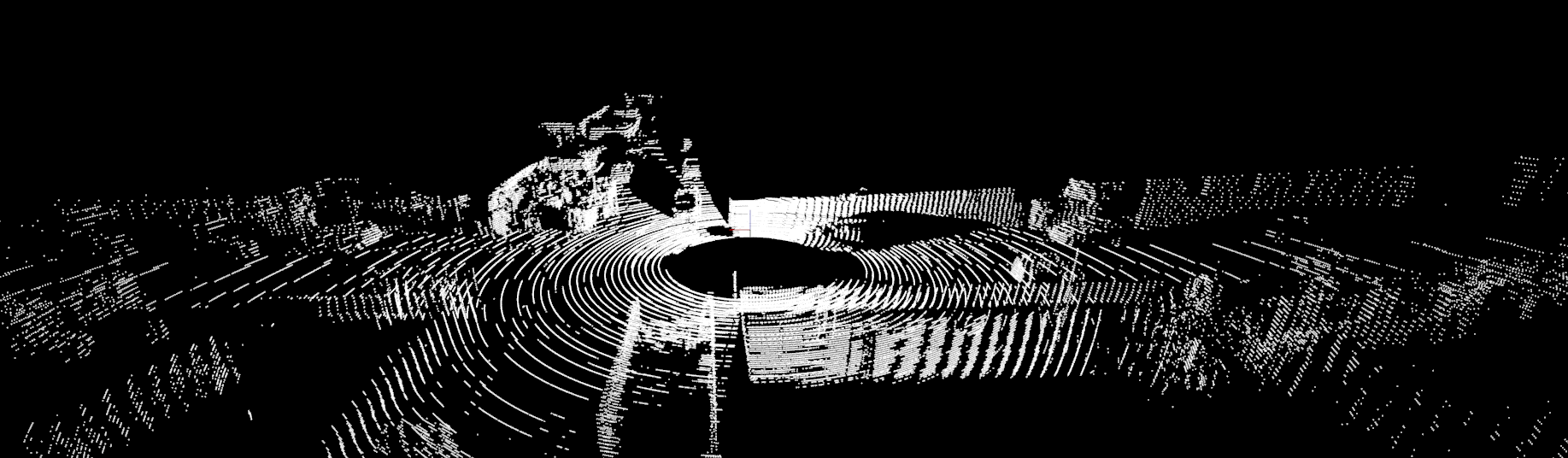}
    \caption{\small Point cloud from 2048$\times$64 RI.}
    \label{fig:ss2048}
  \end{subfigure}\hspace{0.02\textwidth}
  \begin{subfigure}{0.38\textwidth}
    \includegraphics[width=\textwidth]{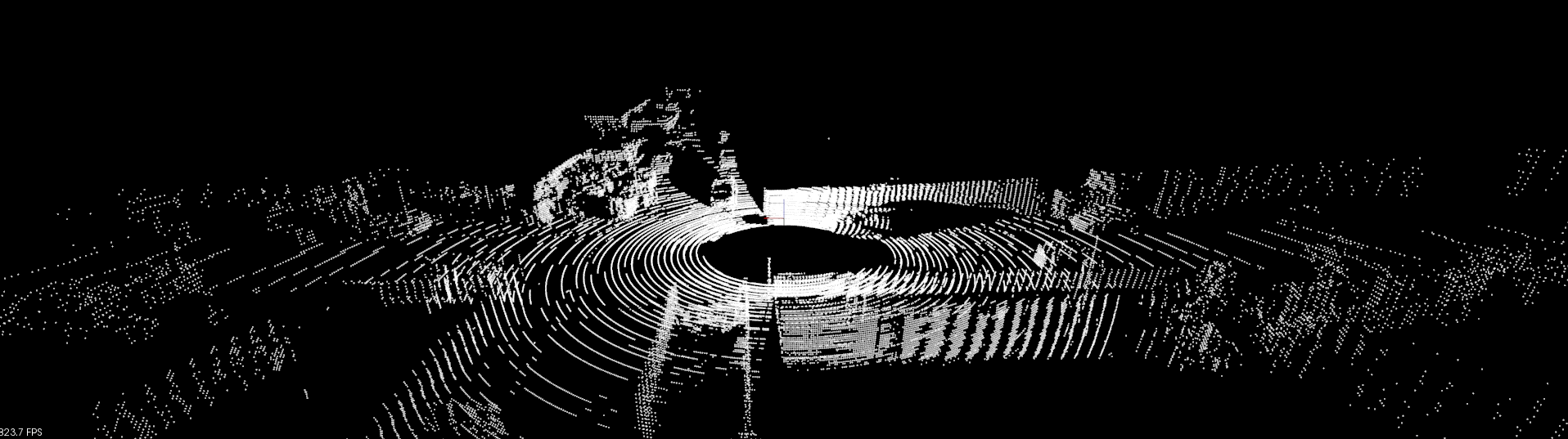}
    \caption{\small Point cloud from 1024$\times$64 RI.}
    \label{fig:ss1024}
  \end{subfigure}

  \begin{subfigure}{0.38\textwidth}
    \includegraphics[width=\textwidth]{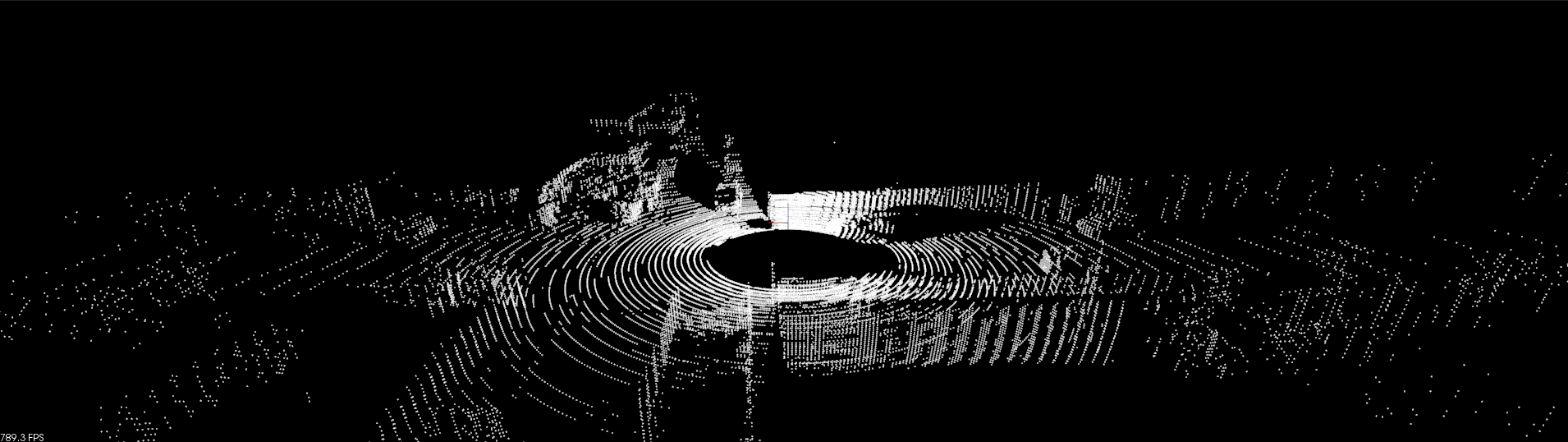}
    \caption{\small Point cloud from 512$\times$64 RI.}
    \label{fig:ss512}
  \end{subfigure}\hspace{0.02\textwidth}
  \begin{subfigure}{0.38\textwidth}
    \includegraphics[width=\textwidth]{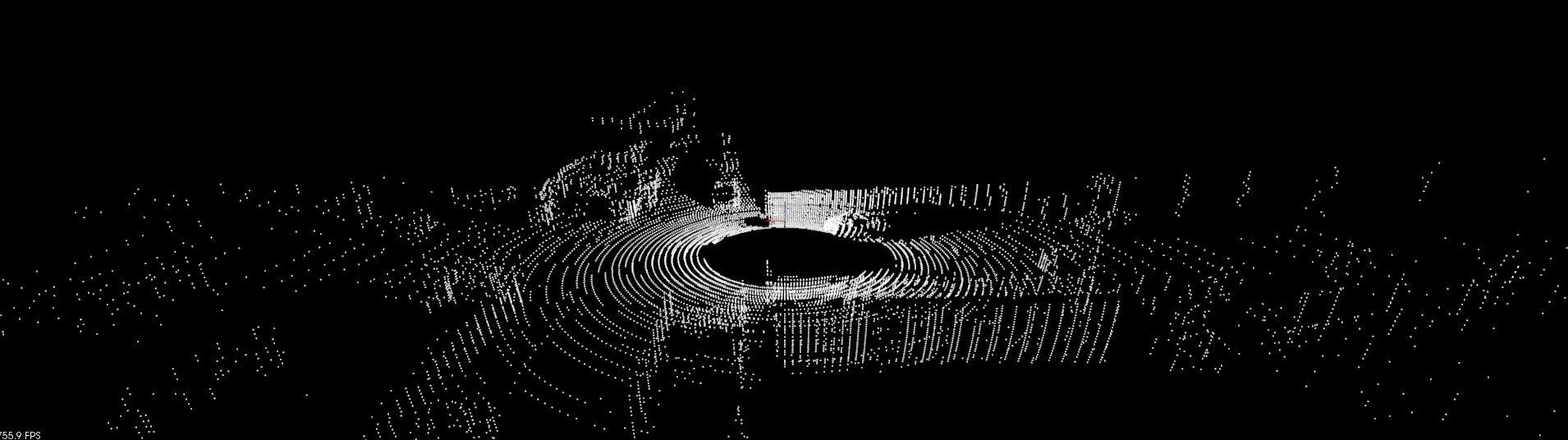}
    \caption{\small Point cloud from 256$\times$64 RI.}
    \label{fig:ss256}
  \end{subfigure}
  \caption {\small The visualizations of the point clouds reconstructed from the subsampled RIs of Figure~\ref{fig:rawpcrep} corresponding to 4500$\times$64 RI.}
  \label{fig:subsample}
\end{figure}

\subsection{RI Quantization and Subsampling}
RI has been used for losslessly mapping LiDAR point clouds to 2D
frames, and previous work only applies quantization of bit-per-point (bpp)~\cite{tu2019real, tu2019point, tu2016compressing, ahn2014large, feng2020real, houshiar20153d}.
In these prior works, the main objective is to maximize the compression efficiency while maintaining the point cloud quality as high as possible.
However, we argue that there are more optimization opportunities with lossy RI to decrease not only the data size but the downstream compression tasks' complexities.
Specifically, the RI resolution is determined by the sensor's precisions as mentioned in Section~\ref{sec:ir}, and the subsampling of point clouds can be done by adjusting the precision parameters; the 3D points are coarsely mapped to a 2D frame.
Figure~\ref{fig:subsample} shows the visualizations of reconstructed point clouds from the subsampled RIs.
From the raw point cloud of Figure~\ref{fig:rawpcrep}, we reduced the precision parameters to map it to the RIs of four different lower resolutions.
Even with the lowest subsampled RI of 16 KB with 8 bpp, the shapes of scanned objects are recognizable.

While the subsampled and quantized RI has advantages for data reduction and compression with lower latencies, it would affect the performance of the perception tasks.
So, a quality metric for the point clouds from lossy RIs needs to reflect both the quantization and subsampling errors.
The currently used metrics, PSNR and CD, reflect the quantization error well, but the sampling error of lossy RIs is not represented effectively as these metrics are defined with the point-to-point distances between point clouds (see Section~\ref{sec:metric} for more details).

Table~\ref{tab:oldmetrics} shows PSNR, CD, and sampling error (SE) of the point clouds from four RI resolutions.
In the results, the changes of PSNR and CD exhibit different trends from SE because SE is about the number of lost points from the original point cloud (the entropy-wise quality) while PSNR and CD are with the distances of the closest point pairs between two point clouds (the point-wise quality).

The current metrics’ issue is they only count the point-to-point distances, and each point distance is calculated by finding the nearest point in the comparing point cloud. So, when the point clouds have different numbers of points, they are limited to represent this difference in the total number of points in the point clouds.
To address the limitations of the existing metrics, we propose a unified metric for both the point-wise quality and the information amount to measure quantitatively the impacts on the downstream perceptions from lossy RIs in Section~\ref{sec:metric}.

\begin{figure}[]
  \centering
  \includegraphics[width=\linewidth]{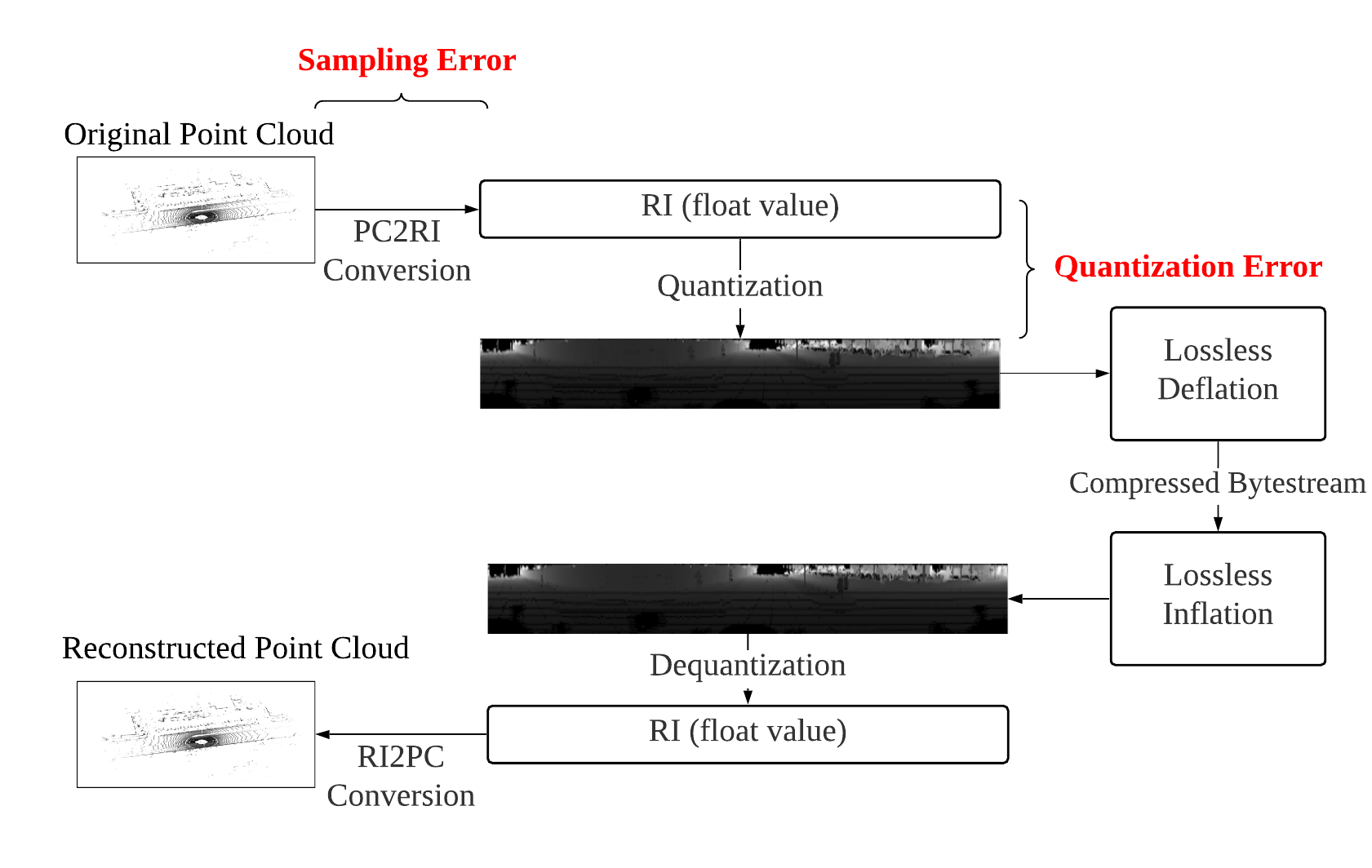}
  \caption{\small FLiCR compression steps with lossy RIs of subsampling and quantization and lossless compression algorithms.}
  \label{fig:rioverview}
\end{figure}

\subsection{Lossless Compression with Lossy RIs}

\begin{figure*}
  \centering
  \begin{subfigure}[t]{0.38\textwidth}
    \includegraphics[width=\textwidth]{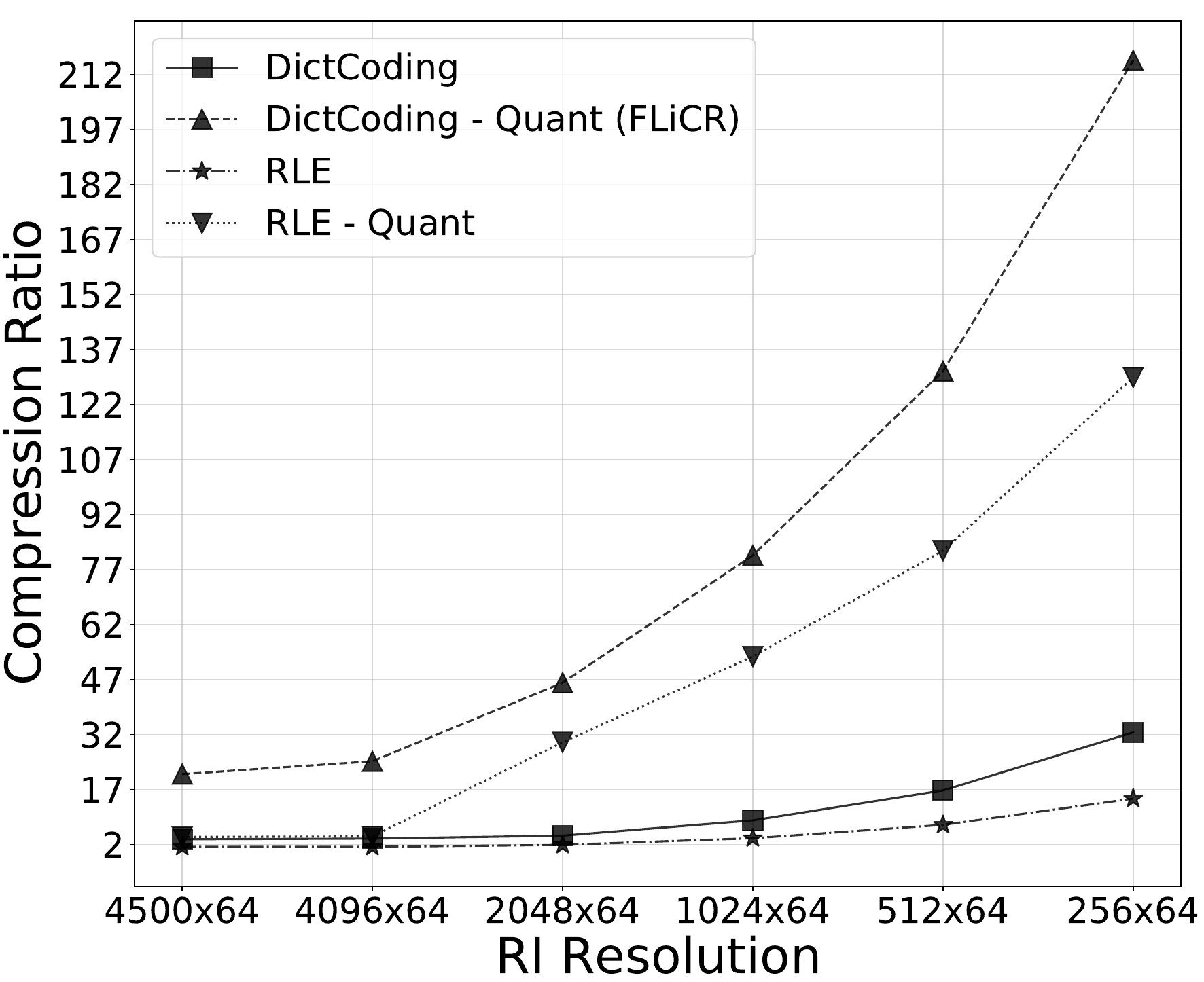}
    \caption{\small The compression ratios with and without quantization in different RI resolutions.}
    \label{fig:losslesscr}
  \end{subfigure}\hspace{0.08\textwidth}
  \begin{subfigure}[t]{0.38\textwidth}
    \includegraphics[width=\textwidth]{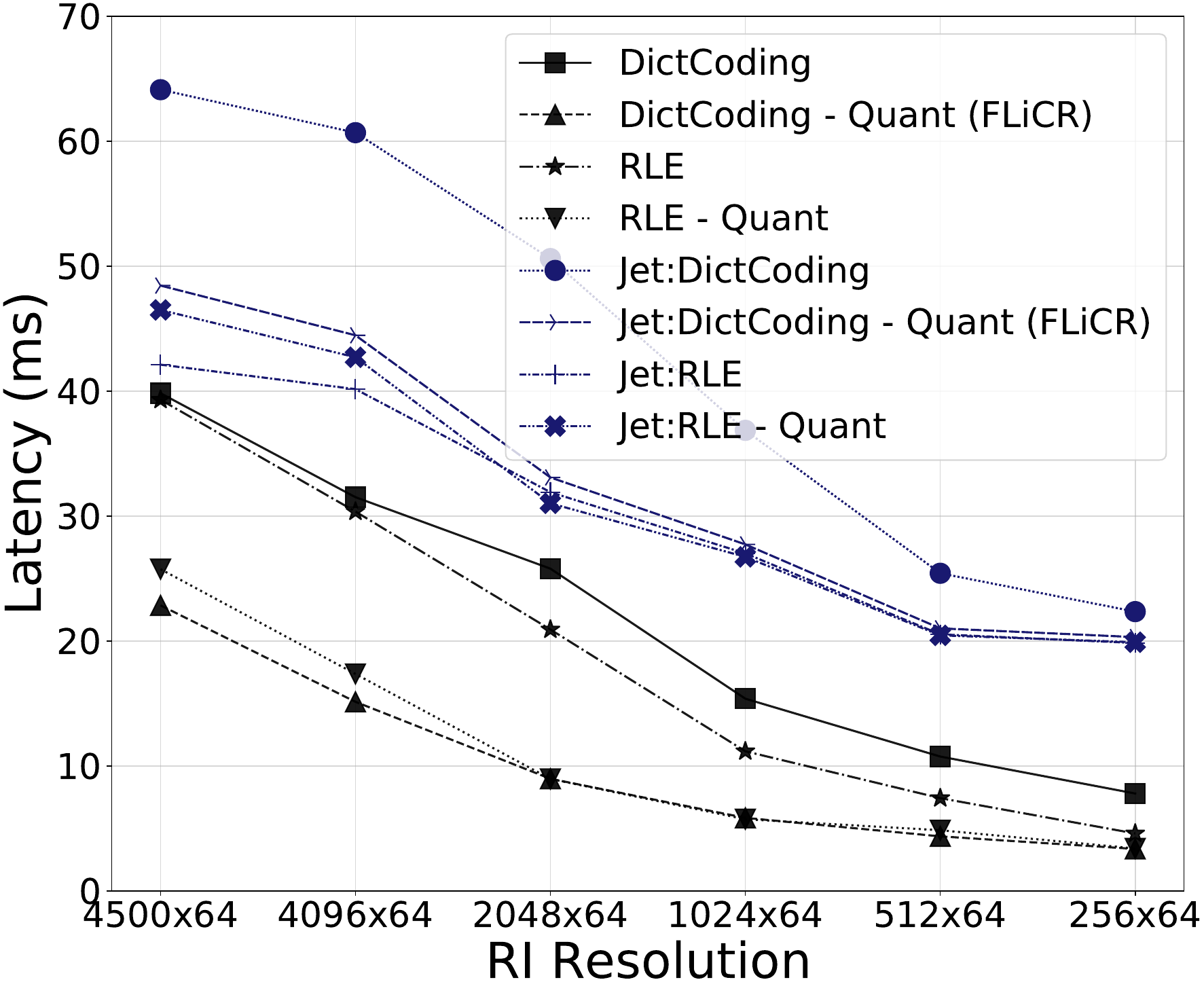}
    \caption{\small The end-to-end latencies with and without quantization in different RI resolutions.}
    \label{fig:losslesslat}
  \end{subfigure}

  \begin{subfigure}[t]{0.38\textwidth}
    \includegraphics[width=\textwidth]{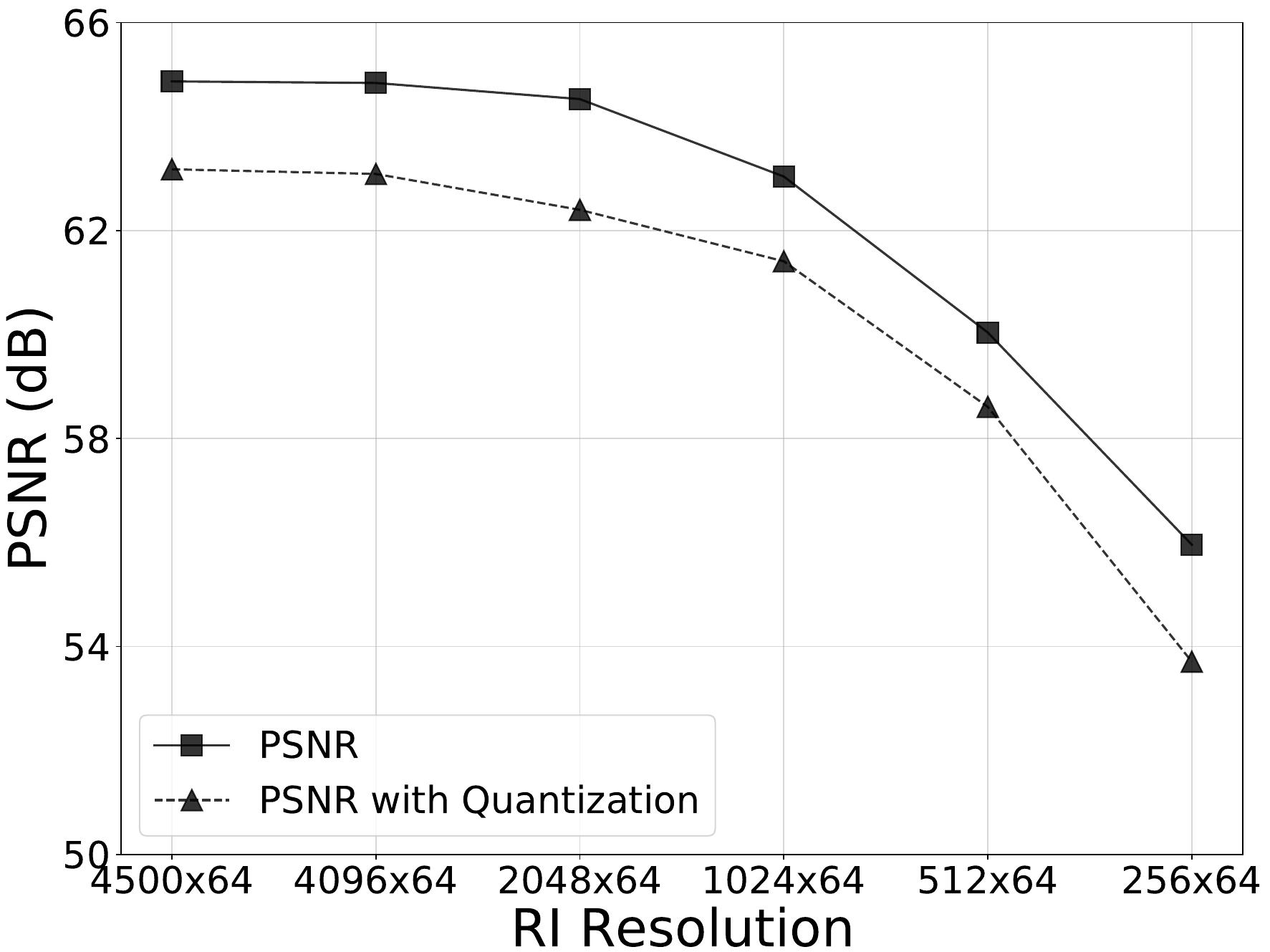}
    \caption{\small The PSNR results with and without quantization in different RI resolutions.}
    \label{fig:losslesspsnr}
  \end{subfigure}\hspace{0.08\textwidth}
  \begin{subfigure}[t]{0.38\textwidth}
    \includegraphics[width=\textwidth]{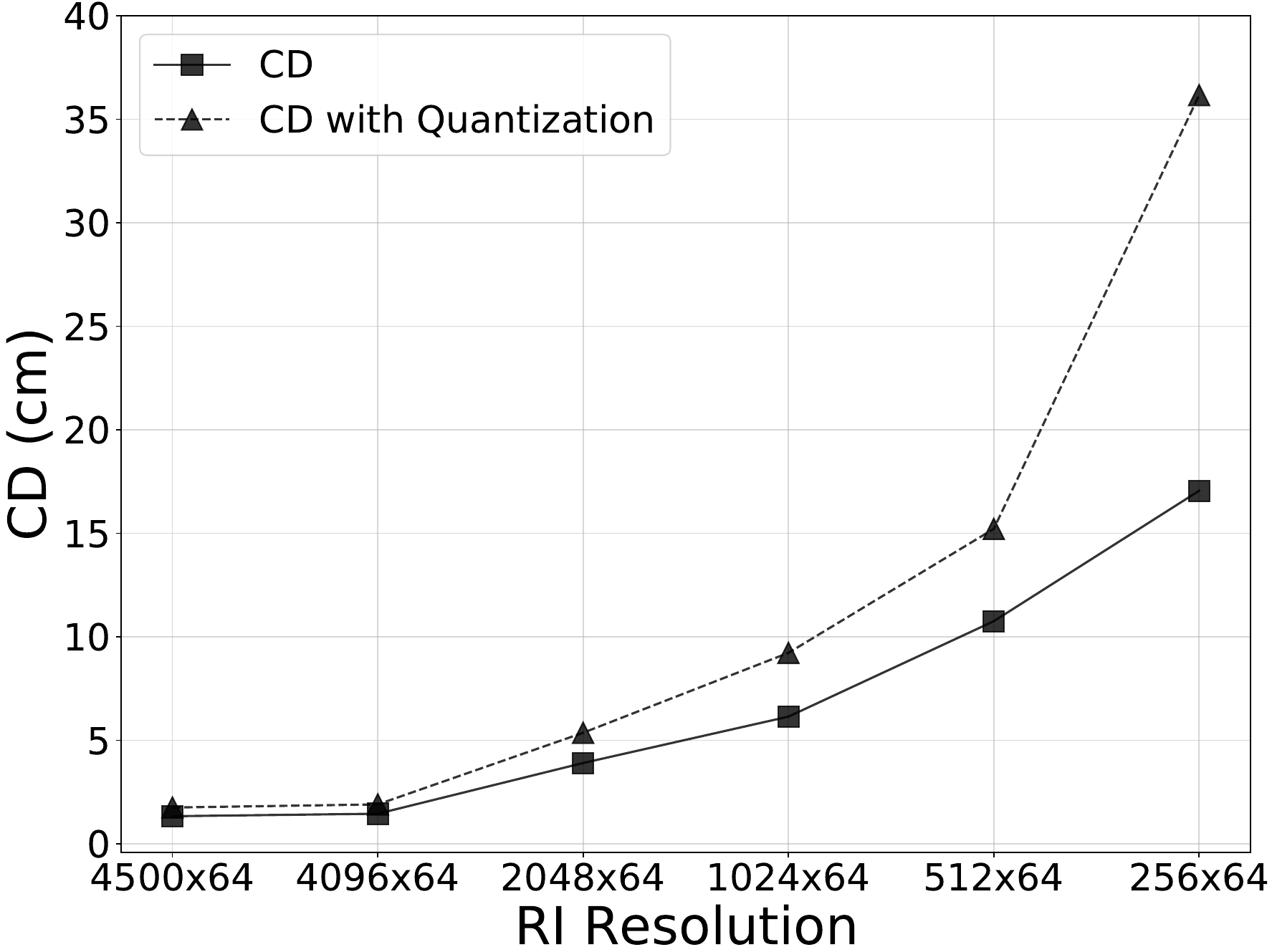}
    \caption{\small The CD results with and without quantization in different RI resolutions.}
    \label{fig:losslesscd}
  \end{subfigure}

  \caption {\small The quantization and subsampling impacts on the lossless bytestream compressions.}
  \label{fig:losslessours}
\end{figure*}

As shown in Section~\ref{sec:issuecodecs}, the application of lossy
video codecs to RIs results in lower compression efficiency or can distort the point clouds in the 3D space.
The previous RI compression methods apply the image compression algorithm at lower efficiencies or propose effective lossless RI compression algorithms via spatial and temporal optimizations~\cite{ahn2014large, feng2020real, houshiar20153d, tu2016compressing, tu2019point, tu2019real}.
%However, they only leverage the quantization of bit precisions with lossless RI mapping and their complex algorithms have downsides for low-latency and lightweight while showing high compression ratios.
However, they partially leverage the opportunities of lossy RIs only
with  bit quantization, and their complex algorithms have downsides
% low-latency and lightweight
in terms of latency and overheads, while showing high compression ratios.
For satisfying the low-latency, lightweight, and efficiency requirements, we use the existing bytestream compression algorithm, dictionary coding, and enhance its efficiency by fully leveraging the RI quantization and subsampling.

Dictionary coding is a lossless compression algorithm for bytestreams
and deflates the bytestream by replacing the repeating patterns with shorter references.
Dictionary coding algorithms have been extensively studied with corpus
text data, and they are with simple bit/byte operations and lower
computation complexities in terms of the space-time tradeoff~\cite{shanmugasundaram2011comparative}.
So, they provide the benefits of being lightweight and low-latency, with simple operations and do not distort point clouds unexpectedly.
Even with such advantages, the direct application of bytestream compressions to the raw point cloud and unquantized RIs of floating values is inefficient in terms of the compression ratio (RLE and Dict Coding in Table~\ref{tab:expcc} and Figure~\ref{fig:losslesscr}).

To improve the efficiency, we fully utilize both quantization and subsampling.
The underlying assumption of our approach is dictionary coding uses the repeating features in a bytestream and there is a higher probability of recurring patterns when limiting the representation space of quantized RIs.
The compression pipeline is shown in Figure~\ref{fig:rioverview}.
FLiCR with dictionary coding has similarity to previous RI-based works in terms of leveraging local spatial features, but it is more advantageous in meeting the requirements with its simplicity.
Explicitly, compared to the recent RI-based compression (RT-ST~\cite{feng2020real} in Table~\ref{tab:expcc}), FLiCR shows lower encoding and decoding latencies and energy usage, with the higher compression ratio, as shown in Table~\ref{tab:oursvsdraco}.

Among the dictionary coding algorithms, we use LZ77~\cite{ziv1977universal} and compare it with RLE.
We measure the efficiency improvement and quality reduction by the quantization and subsampling with LZ77 and RLE, and Figure~\ref{fig:losslessours} shows the results of different resolutions of RIs quantized by 8 bpp.
While the end-to-end latencies of the whole compression pipeline can
decrease only with subsampling (see Figure~\ref{fig:losslesslat}),
both quantization and subsampling are required to improve the
compression ratios effectively, as shown in Figure~\ref{fig:losslesscr}.
Then, dictionary coding shows  larger growth than RLE, and these results support our assumption about the performance improvement of dictionary coding with lossy RIs.

Although FLiCR achieves compression efficiency and reduced latency, it
is at the cost of degradation of the point cloud quality by the quantization and subsampling errors, as shown in Figures~\ref{fig:losslesspsnr} and~\ref{fig:losslesscd}.
Since the reduced point cloud quality can have an impact on the downstream perceptions, we evaluate FLiCR with the state-of-the-art LiDAR perceptions and analyze the errors' impacts in Section~\ref{sec:eval}.

\begin{figure}[htbp]
  \centering
  \begin{subfigure}{0.4\textwidth}
    \includegraphics[width=\textwidth]{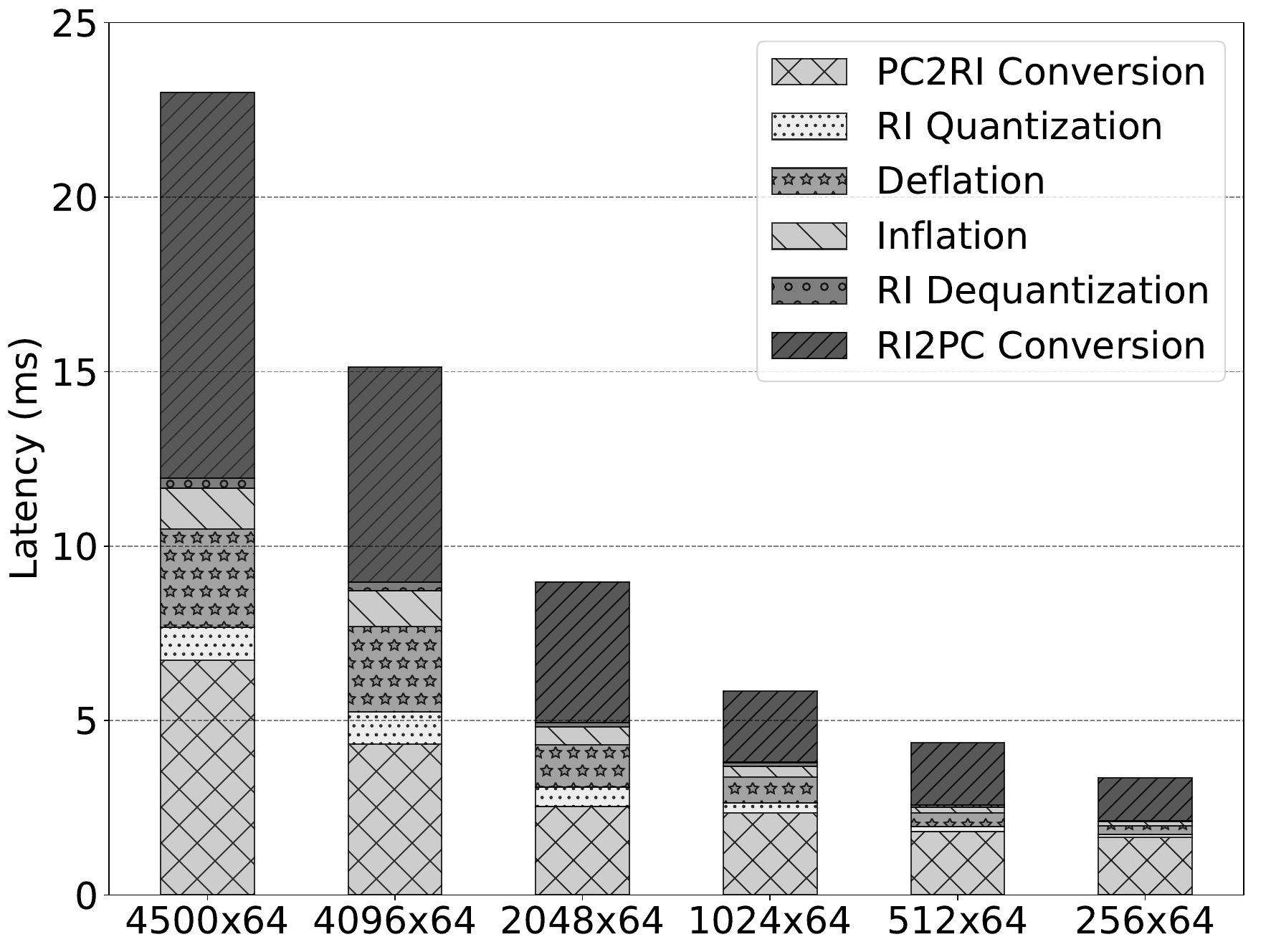}
    \caption{\small The latency breakdown of our method on desktop.}
    \label{fig:latbreakdesk}
  \end{subfigure}\hspace{0.01\textwidth}
  \begin{subfigure}{0.4\textwidth}
    \includegraphics[width=\textwidth]{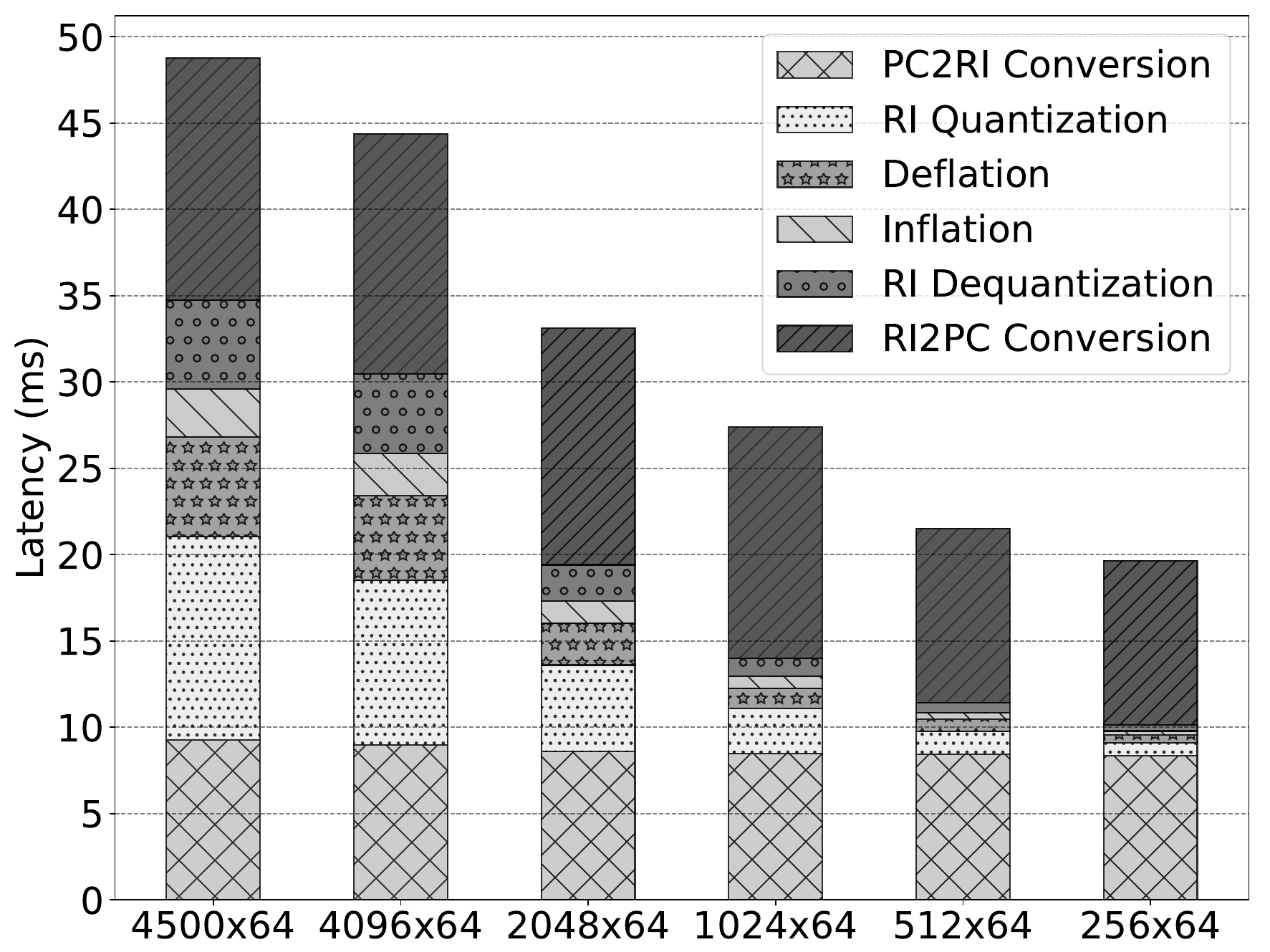}
    \caption{\small The latency breakdown of our method on Jetson AGX.}
    \label{fig:latbreakjet}
  \end{subfigure}

  \caption {\small The end-to-end latency breakdowns of FLiCR.}
  \label{fig:latbreak}
\end{figure}

Figure~\ref{fig:latbreak} shows the latency breakdowns of FLiCR on our testbed.
In the case where Jetson is a mobile client and the desktop is a server, the end-to-end latency is $\sim$39 ms ($\sim$60\% of Google Draco~\cite{draco}) even with the highest RI resolution; it takes 27 ms for client encoding and 12 ms for server decoding.
With the 256$\times$64 resolution, the end-to-end latency is $\sim$10 ms which is $\sim$16\% of Draco.
Since the large portion of the end-to-end latency is the conversion time between the point cloud and RI, the end-to-end latencies can be largely reduced if a device has a dedicated hardware logic for the RI conversion.
As the RI resolution gets lower, the quantization and compression latencies decrease.
These results show FLiCR fully leverages the synergistic effect by quantization and subsampling for the bytestream compressions.

%------------------------------------------------------------------------------
\section{\lowercase{e}PSNR: Quality Metric for L\lowercase{i}DAR Point Clouds}
\label{sec:metric}
%------------------------------------------------------------------------------

The errors in the RI conversion process can affect the performance of LiDAR perceptions.
PSNR and CD (or RMSE) have been broadly used as the quality metrics of
3D point clouds~\cite{biswas2020muscle, huang2020octsqueeze, tu2016compressing, que2021voxelcontext, feng2020real, tu2019point, tu2019real}, but by definition they are point-wise quality metrics and do not reflect the point loss effectively.
In the context of our approach with lossy RIs, we argue a metric for both point-wise quality and overall information amount is essential.

\begin{equation}\label{eq:2}
Dist(p, C) = \min_{{p_c}}\left( (p_c-p)^2 \right)
\end{equation}

\begin{equation}\label{eq:2-2}
MSE(C_1, C_2) = \frac{1}{\left\| C_2 \right\|}\sum_{i=0}^{\left\| C_2 \right\|-1} \left\{ Dist(p_{c_2}, C_1) \right\}
\end{equation}

Both PSNR and CD use the mean squared error (MSE) of the point-wise distances between two point clouds.
When $C_1$ is the original point cloud and $C_2$ is the reconstructed point cloud, the distance between a point in $C_2$ and the corresponding point in $C_1$ is calculated by Equation~\ref{eq:2}.
So, the corresponding point in $C_1$ is of the shortest distance to the point in $C_2$.
Then, MSE between two point clouds is defined by Equation~\ref{eq:2-2}.

\begin{equation}\label{eq:3}
  \scalebox{.9}{$CD(C_{orig}, C_{comp}) = MSE(C_{orig}, C_{comp}) + MSE(C_{comp}, C_{orig})$}
\end{equation}

\begin{equation}\label{eq:4}
  \scalebox{.9}{$PSNR(C_{orig}, C_{comp}) = 10\ log\left( \frac{Max^2}{MSE(C_{orig}, C_{comp})} \right)$}
\end{equation}

Then, PSNR and CD are defined as Equation~\ref{eq:3} and~\ref{eq:4}.
CD is the sum of reciprocal MSEs between two point clouds, and PSNR is the ratio of the peak LiDAR sensor range to MSE.
As indicated by their definitions, these metrics are based on MSE and are determined by the point-to-point distance.
They natively represent the quality loss by the quantization error, but the subsampling error is not effectively represented even if the impact of SE is shown mildly as some nearest points can be lost in the reconstructed point cloud.
Specifically, in the current metrics, it is possible to get a high-quality result even with a few points of  small distances to a point cloud of  large number of points.
It is caused by the unstructured nature of the LiDAR point cloud; there is no point-to-point correspondence between LiDAR point clouds having different numbers of points.
In the case of the normal images, the total number of pixels is fixed without SE and the point-wise metrics work well.

\begin{equation}\label{eq:5}
SE = \frac{\left\| C_{orig} - C_{comp} \right\|}{\left\| C_{orig} \right\|},\ 0 \le SE\le 1
\end{equation}

Based on our observation, we argue it is inappropriate to use the metrics representing only the point-wise quality for LiDAR point clouds.
To address the limitation of the current metrics, we propose a new
single-number metric, {\bf entropy-reflecting PSNR (ePSNR)}, by extending PSNR.
ePSNR is designed to indicate both the point-wise and entropy-wise quality of a point cloud.

SE is related to the total information (entropy) loss in a point cloud
because it is the percent of the lost points, as in Equation~\ref{eq:5}.
One naive way of making PSNR reflect the entropy is to multiply $1-SE$ to PSNR with the assumption that the entropy is $1-SE$ and SE is exactly the same with the actual entropy loss, $\mathscr{L}_{SE}$.
However, instead of the naive way, we extend PSNR by estimating $\mathscr{L}_{SE}$.
Our underlying assumption is $\mathscr{L}_{SE}$ is not exactly the same with SE and follows the exponential distribution as Equation~\ref{eq:6}.
The intuition for this assumption is that SE can have minimal impacts on the downstream perceptions as far as the total amount of necessary information is preserved for the perception algorithms.
It means there would be a knee of the curve in the graph of the entropy function.

\begin{equation}\label{eq:6}
  Assumption:\quad\mathscr{L}_{SE} \sim \mathcal{\text{exp}(\beta)}
\end{equation}

When $\mathscr{L}_{SE}$ follows the exponential distribution, the entropy function $\mathcal{F}(SE)$ can be defined with the cummulative distribution function (CDF) of the exponential distribution as Equation~\ref{eq:7}.
This entropy function is a probability function estimating the actual entropy of the remaining points in a point cloud with the given SE.

\begin{equation}\label{eq:7}
\mathcal{F}(SE) = \mathrm{P}(\mathrm{E}>x) = e^{-\frac{x}{\beta}}\quad where\ x = 1-SE \\
\end{equation}

With our entropy function, ePSNR is defined as Equation~\ref{eq:8}.
Since it is based on PSNR, the point-wise quality with the quantization error is represented while reflecting the entropy with the given SE.
When SE is small, ePSNR would be almost same with the original PSNR,
but would start to decrease exponentially when SE gets larger, by its definition.
ePSNR has two parameters: $\alpha$ as a derivative adjusting factor to prevent too steep or shallow distribution and $\beta$ of the exponential distribution.

\begin{equation}\label{eq:8}
\begin{gathered}
\scalebox{.9}{$ePSNR(C_{orig}, C_{comp}) =  PSNR\times \left\{ 1 - (SE \times (\mathcal{F}(SE)+\alpha)) \right\},$}
  \\ \scalebox{.9}{$0\le \mathcal{F}(SE)+ \alpha \le 1$}
\end{gathered}
\end{equation}

%------------------------------------------------------------------------------
\section{Evaluation}
\label{sec:eval}
%------------------------------------------------------------------------------
The goal of this section is to demonstrate that our approach appropriately meets the requirements of the LiDAR point cloud compression for enabling edge-assisted online perceptions.
We compare FLiCR with several existing compression method.
Since FLiCR affects the quality of the point clouds, its impact is evaluated with the state-of-the-art LiDAR perception algorithms for 3D objection detection and LiDAR odometry and mapping (LOAM).
We also evaluate ePSNR and demonstrate its effectiveness compared to PSNR and the naive way of combining PSNR and SE.
All experiments are done by using the LiDAR point clouds of the KITTI dataset~\cite{geiger2013vision}, which are captured from Velodyne HDL-64E~\cite{hdl64}.

\subsection{Experimental Testbed}
\label{sec:subsec_exp}
The testbed consists of two machines, an NVIDIA Jetson AGX Xavier and a high-end desktop.
The Jetson has ARMv8 CPU and 32 GB memory, and
we set its power mode as 15W.
The desktop has Intel Core i7-10700, 32 GB memory, and NVIDIA RTX 2070 GPU.
Both run Ubuntu 18.04, and the energy usage is measured with \texttt{perf} on desktop and \texttt{tegrastats} on Jetson.

\subsection{FLiCR Benchmark}
As shown in Table~\ref{tab:expcc}, Google Draco~\cite{draco} is the most suitable compression method for online perceptions in terms of compression ratio, latency, and energy usage.
So, we compare FLiCR with different RI resolutions to Draco.
We benchmark FLiCR and measure the compression ratio, SE, PSNR, ePSNR ($\alpha=-0.15$, $\beta=0.5$), latencies, and energy usage.
The compression level parameter of Draco is set as 10 which is maximum.
For FLiCR, each RI is quantized by 8 bpp.
The benchmark results are shown in Table~\ref{tab:oursvsdraco}.

% Table - Draco vs FLiCR 4500, 4096, 2048, 1024, 512, 256
\begin{table*}[]
  \caption{\label{tab:oursvsdraco} \small The comparison between Google Draco and FLiCR of different RI resolutions.}
  \begin{tabularx}{\textwidth}{{|>{\raggedright}X
                                |>{\centering}c
                                |>{\centering}c
                                |>{\centering\arraybackslash}c
                                |>{\centering\arraybackslash}c
                                |>{\centering\arraybackslash}c
                                |>{\centering\arraybackslash}c
                                |>{\centering\arraybackslash}c|}}
  \hhline{========}
                         & Google Draco   & \specialcell{FLiCR \\ 4500$\times$64}          & \specialcell{FLiCR \\ 4096$\times$64}         &  \specialcell{FLiCR \\ 2048$\times$64}         & \specialcell{FLiCR \\ 1024$\times$64}         & \specialcell{FLiCR \\ 512$\times$64}          & \specialcell{FLiCR \\ 256$\times$64} \\ \hline
    Compression Ratio    & 17.05          & 21.26                      & \textbf{24.75}                     & 46.18                      & 80.88                     & 131.13                    & 215.85           \\ \hline
    SE                   & \textbf{6.7\%} & 8.4\%                      & 9.1\%                     & 21\%                       & 58.8\%                    & 78.9\%                    & 89.2\%           \\ \hline
    PSNR (dB)            & \textbf{67.29} & 63.18                      & 63.09                     & 62.4                       & 61.41                     & 58.61                     & 53.71            \\ \hline
    ePSNR (dB)           & \textbf{67.27} & 63.13                      & 63.01                     & 61.64                      & 51.38                     & 35.4                      & 22.29            \\ \hline
    Enc Time (ms)        & 21.1 (48.4)    & 10.48 (26.83)              & \textbf{7.69} (\textbf{23.41})              & 4.3 (16.03)                & 3.37 (12.26)              & 2.35 (10.46)              & 1.99 (9.54)      \\ \hline
    Dec Time (ms)        & 9.44 (18.6)    & 12.52 (21.94)              & \textbf{7.44} (\textbf{20.97})              & 4.67 (17.09)               & 2.47 (15.13)              & 2.01 (11.06)              & 1.36 (10.11)     \\ \hline
    Enc Energy Usage (J) & 0.83 (0.14)    & 0.36  (0.09)               & \textbf{0.27} (\textbf{0.07})               & 0.16 (0.04)                & 0.13 (0.03)               & 0.09 (0.03)               & 0.07 (0.02)      \\ \hline
    Dec Energy Usage (J) & 0.36 (0.05)    & 0.48  (0.05)               & \textbf{0.3}  (\textbf{0.05})               & 0.19 (0.04)                & 0.13 (0.04)               & 0.09 (0.03)               & 0.08 (0.02)      \\
  \hhline{========}
  \end{tabularx}
\end{table*}

FLiCR achieves higher compression ratios across all resolutions than Draco, and it is $\sim$25\% higher even with the highest resolution.
As highlighted in Table~\ref{tab:oursvsdraco}, FLiCR with little subsampling starts to outperform Draco in the compression ratio, latency, and energy usage.
One characteristic of Draco is its encoding process takes longer and uses more energy than decoding while with FLiCR it is similar in most cases.
Considering the use case of edge-assisted perceptions, it is the client who encodes and sends data, and the server receives and decodes the encoded data to feed it to perception modules.
When the client is Jetson and the server is the desktop in our testbed, Draco introduces  $\sim$60 ms end-to-end latency, and with FLiCR it ranges from $\sim$11 ms for 256$\times$64 to $\sim$40 ms for 4500$\times$64.
Therefore, FLiCR is more advantageous for the latency-performance tradeoff of online perceptions  for commodity mobile devices, given its higher compression efficiency in terms of compression ratios and energy usage.

To satisfy the aforementioned requirements, we compromise the quality of point clouds with lossy RIs.
Since Draco quantizes each point as 11 bpp, it shows a higher PSNR, and ePSNR is almost same with PSNR as it has  small SE.
One issue about SE is that the highest resolution RIs have $\sim$8\% SE even though its resolution is with the maximum sensor precision of HDL64 specified in the spec sheet~\cite{hdl64}.
We presume this SE is caused by the sensor's measurement error and noise as the LiDAR point clouds are captured by running cars.
In Table~\ref{tab:oursvsdraco}, PSNR barely changes with 1024$\times$64 RIs which have 58.8\% SE while ePSNR reflects it.
Since the reduced quality affects the performance of downstream perceptions, we also evaluate our compression and metric with the state-of-the-art LiDAR perceptions.

\subsection{End-to-end Evaluation}
We evaluate our method and metric with two perception tasks: 3D object detection and LOAM.
For 3D object detection, we use machine learning (ML) models pre-trained with the original point clouds from the KITTI dataset from the Model Zoo of OpenPCDet~\cite{openpcdet2020}. We use the following models: Part-$A^2$ Net~\cite{shi2020points}, PointPillars~\cite{lang2019pointpillars}, PointRCNN~\cite{shi2019pointrcnn}, PV-RCNN~\cite{shi2020pv}, SECOND~\cite{yan2018second}, Voxel R-CNN~\cite{deng2020voxel}.
These models are trained with 7481 samples, and the testset is 7518 LiDAR scans.
For LOAM~\cite{zhang2014loam}, we use the A-LOAM implementation~\cite{aloam}.
For checking the impacts of RI quantization and subsampling, we generate the LiDAR point cloud dataset reconstructed from different resolution RIs.
Then, we feed our dataset to those perception models.
Since the object detection models are trained with the original LiDAR data and A-LOAM is implemented and tested by using the original dataset, we can quantitatively measure the impacts of lossy RIs in FLiCR on the perception performance.

\begin{table*}[]
  \caption{\label{tab:exp3dobj} \small The 3D object detection performances with different IoU threshold and reconstructed point clouds from the RIs. The number in each cell is the recall for the detected objects in the scene.}
  \begin{tabularx}{\textwidth}{{|>{\raggedright}X
                                ||>{\centering}c
                                |>{\centering}c
                                |>{\centering\arraybackslash}c
                                |>{\centering\arraybackslash}c
                                |>{\centering\arraybackslash}c
                                |>{\centering\arraybackslash}c
                                |>{\centering\arraybackslash}c
                                |>{\centering\arraybackslash}c|}}
  \hhline{=========}
    \multicolumn{2}{|l|}{}                                        & Original     & 4500$\times$64 RI & 4096$\times$64 RI  & 2048$\times$64 RI& 1024$\times$64 RI & 512$\times$64 RI & 256$\times$64 RI   \\ \hline
    \multirow{6}{8em}{IoU Threshold \\ \hfil 0.3}          & Part-$A^2$ Net   & 95.1         & 88.4              & 88.3               & 87.9             & 76.2              & 75.5             & 56.1               \\ \cline{2-9}
                                                           & \textbf{PointPillars} & \textbf{94}           & \textbf{92.6}              & \textbf{92.4}               & \textbf{91.5}             & \textbf{82}                & \textbf{76.1}             & \textbf{54.8}               \\ \cline{2-9}
                                                           & PointRCNN    & 89.8         & 70.7              & 71.2               & 71.7             & 68.8              & 62.9             & 47.5               \\ \cline{2-9}
                                                           & PV-RCNN      & 96.8         & 94.6              & 94.4               & 94               & 89.9              & 82.8             & 70.2               \\ \cline{2-9}
                                                           & SECOND       & 94.9         & 92.7              & 92.6               & 92.1             & 88.4              & 79.4             & 60.2               \\ \cline{2-9}
                                                           & Voxel R-CNN  & 95.4         & 93.6              & 93.6               & 93.5             & 89                & 87.5             & 76.1               \\ \hline
    \multirow{6}{8em}{IoU Threshold \\ \hfil 0.5}       & Part-$A^2$ Net  & 91.2         & 82.3              & 82.2               & 81.1             & 71.4              & 63.7             & 41.3               \\ \cline{2-9}
                                                           & \textbf{PointPillars} & \textbf{88.7}         & \textbf{82.7}              & \textbf{82.5}               & \textbf{81}               & \textbf{69.8}              & \textbf{57.8}             & \textbf{30.8}               \\ \cline{2-9}
                                                           & PointRCNN    & 87.1         & 65.5              & 66                 & 66.2             & 63.8              & 57               & 38.5               \\ \cline{2-9}
                                                           & PV-RCNN      & 93.4         & 88.9              & 88.8               & 87.6             & 81.8              & 71.2             & 53                 \\ \cline{2-9}
                                                           & SECOND       & 89.1         & 83.8              & 83.7               & 82.3             & 76.8              & 61.9             & 38                 \\ \cline{2-9}
                                                           & Voxel R-CNN  & 94.9         & 91.1              & 91.1               & 90.5             & 85.2              & 79.4             & 58.4               \\ \hline
    \multirow{6}{8em}{IoU Threshold \\ \hfil 0.7}       & Part-$A^2$ Net  & 73.6         & 59.9              & 59.8               & 57.4             & 46                & 38.1             & 21.6               \\ \cline{2-9}
                                                           & \textbf{PointPillars} & \textbf{63.9}         & \textbf{49.6}              & \textbf{49.3}               & \textbf{46}               & \textbf{33.4}              & \textbf{18.4}             & \textbf{5.5}               \\ \cline{2-9}
                                                           & PointRCNN    & 73.3         & 46.8              & 47.3               & 46.9             & 43.9              & 36.4             & 20.8               \\ \cline{2-9}
                                                           & PV-RCNN      & 75.9         & 60.6              & 60.4               & 57.4             & 49                & 33.3             & 16.7               \\ \cline{2-9}
                                                           & SECOND       & 66.5         & 52.4              & 52.3               & 49.1             & 41.5              & 26.3             & 10.3               \\ \cline{2-9}
                                                           & Voxel R-CNN  & 84.6         & 67.9              & 67.9               & 64               & 54.1              & 37.1             & 16.4               \\
  \hhline{=========}
  \end{tabularx}
\end{table*}

\noindent{\textbf{3D Object Detection.\quad}}
3D object detection is the task of detecting objects from 3D point clouds.
Each algorithm of the models we use has a different network architecture, but there is a commonality between them: a backbone network extracts features from the point clouds and the extracted features are used by the regional proposal networks (RPN).
As the backbone networks of these models, PointNet++~\cite{qi2017pointnet++} is used to extract the point-level features.
For the voxel-level features, the voxel feature encoder (VFE) layer and 3D sparse convolutional networks~\cite{3DSemanticSegmentationWithSubmanifoldSparseConvNet} are used.
When each model produces the region proposals of the detected objects, they are compared with the region of the ground truth objects.
The result recall is determined by the detected objects corresponding to the ground truth object with the IoU threshold.

Table~\ref{tab:exp3dobj} shows the recall for the detected objects of the models.
With the highlighted example of PointPillars, the performance reductions between the original and 4500$\times$64 RIs show the impact of the quantization error.
In the case of  IoU threshold 0.7, it shows $\sim$23\% performance reduction while it is $\sim$2\% with threshold 0.3.
This shows the results of  higher IoU thresholds are more sensitive to the quantization error, and 3D object detection with a higher IoU threshold requires  input point clouds of almost the same quality as the original training data.

The results across the different resolutions show the impacts of SE.
One noticeable thing is the performance results decrease little with 2048$\times$64 RIs compared to 4500$\times$64 RIs, and this trend is for all IoU thresholds.
These results support our assumption for ePSNR in Section~\ref{sec:metric}; there is a knee of the curve in the entropy loss by SE.
Moreover, Table~\ref{tab:oursvsdraco} shows the ePSNR results drop drastically from 1024$\times$64 RIs as does the performance of the 3D objection detection models.

\begin{figure}[]
  \centering
  \includegraphics[width=0.9\linewidth]{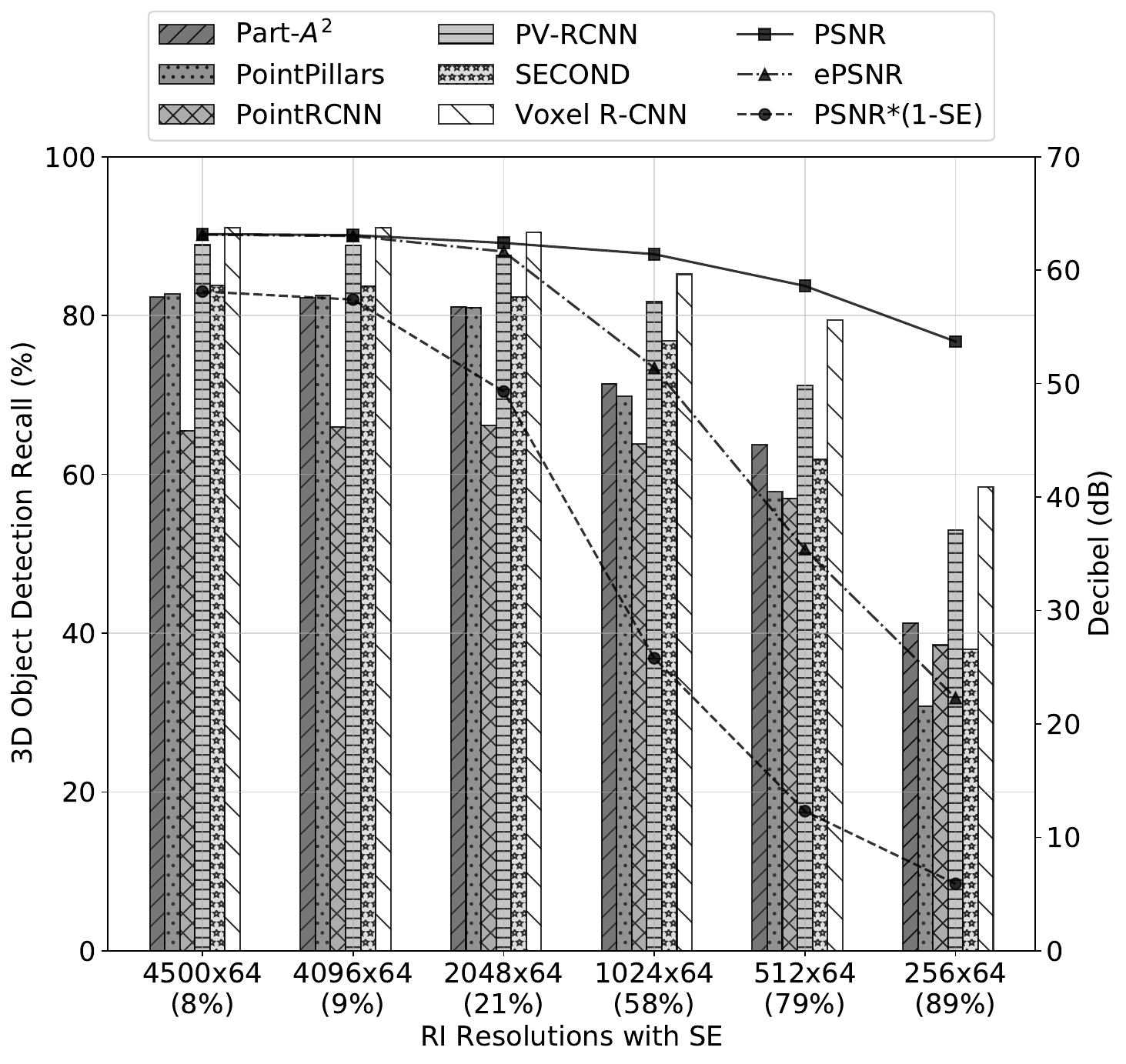}
  \caption{\small The results of PSNR, ePSNR, and the naively-entropy-reflecting PSNR with the 3D object detection results of the IoU threshold 0.5 for the models.}
  \label{fig:epsnrres}
\end{figure}

Figure~\ref{fig:epsnrres} shows the object detection recall values, and PSNR, ePSNR, and the naive way of making PSNR capture entropy, PSNR$\times(1-SE)$, as described in Section~\ref{sec:metric}.
The IoU threshold is 0.5 for all detection models, and the parameters of ePSNR are $\alpha$ (-0.15) and $\beta$ (0.5).
For the changes of SE and recalls, PSNR mildly changes across the RI resolutions, and PSNR$\times(1-SE)$ shows 
more drastic decreases compared to the perception results.
On the other hand, ePSNR shows a similar trend with the performance reduction of the perception models.
These results demonstrate the effectiveness of ePSNR with the probability function estimating the actual entropy by using SE, as a single-number metric for the point-wise and entropy-wise qualities of a point cloud.

\noindent{\bf LiDAR Odometry and Mapping.} LOAM (or LiDAR SLAM) is a 3D mapping technique running the odometry, point matching, and registration (mapping) algorithms simultaneously~\cite{zhang2014loam}.
LOAM, and other SLAM algorithms that use different sensors, are widely used in various use cases, including autonomous vehicle, extended reality, and 3D reconstruction, and are one of the key perception tasks.
We evaluate the quality impacts of point clouds reconstructed from different RI resolutions with A-LOAM~\cite{aloam} and the evaluator~\cite{lidarslamevaluator2021}.
The experiments are with a sequence of 1101 LiDAR point clouds from the KITTI dataset.
We show our evaluation results using two metrics: absolute trajectory error (ATE) and relative error (RE).
While ATE calculates the root mean squared errors (RMSE) of position (ATE$_{pos}$) and rotation (ATE$_{rot}$) to the groundtruth, RE measures the relative relations of sub-trajectories in position (RE$_{pos}$) and rotation (RE$_{rot}$)~\cite{zhang2018tutorial}.

\begin{table}[]
  \caption{\small The LOAM averaged results of the error metrics: Position (m) and Rotation (degree).}
  \begin{center}
  \begin{tabular}{ |c|c|c|c|c| }
    \hline
                      & ATE$_{pos}$ & ATE$_{rot}$ & RE$_{pos}$  & RE$_{rot}$  \\ \hline
    Original          & 0.316       & 0.57        & 0.389       & 0.82        \\ \hline
    4500$\times$64 RI & 0.321       & 0.2         & 0.387       & 0.83        \\ \hline
    4096$\times$64 RI & 0.313       & 0.21        & 0.39        & 0.84        \\ \hline
    2048$\times$64 RI & 0.294       & 0.17        & 0.388       & 0.82        \\ \hline
    1024$\times$64 RI & 0.394       & 0.17        & 0.388       & 0.82        \\ \hline
    512$\times$64  RI &\textbf{0.610}       & 0.17        & 0.388       & 0.82        \\ \hline
    256$\times$64  RI &\textbf{0.596}       & 0.2         & 0.387       & 0.82        \\ \hline
  \end{tabular}
  \end{center}
  \label{tab:loammetric}
\end{table}

\begin{figure}[]
  \centering
  \includegraphics[width=0.8\linewidth]{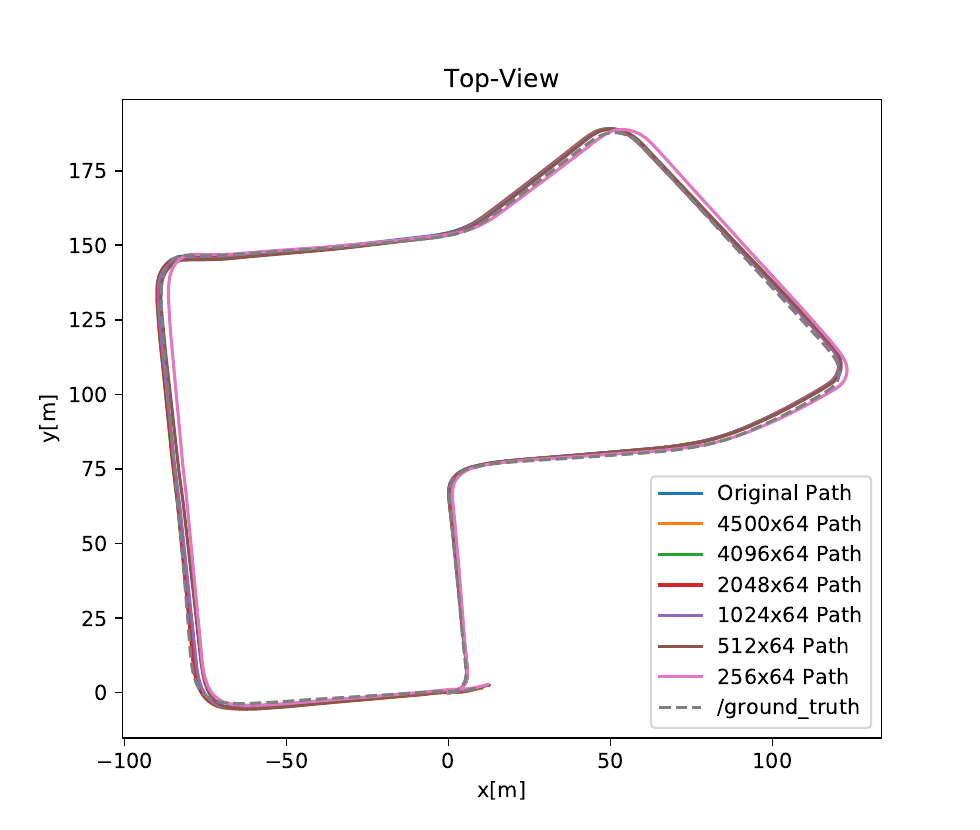}
  \caption{\small The path results of LOAM with point clouds of different RIs.}
  \label{fig:loam}
\end{figure}

Table~\ref{tab:loammetric} shows the evaluation results of A-LOAM with different RIs.
Based on the results, the LOAM algorithm works well even with high quantization and subsampling errors.
Except for the increased ATE$_{pos}$ for 512$\times$64 and 256$\times$64, other results are almost same with the result of the original data.
Moreover, the LOAM paths of all cases are almost identical to each other as shown in Figure~\ref{fig:loam}.
After thorough analysis of the A-LOAM implementation, we find the mapping resolutions of A-LOAM are attributed to these results; the line and plane mapping resolutions of A-LOAM are 0.4 m and 0.8 m~\cite{aloam}.
The increased ATE$_{pos}$ for 512$\times$64 and 256$\times$64 are because the coarser subsampling causes loss of the sparse regions in the scene.
Specifically, in Figure~\ref{fig:subsample}, the points over long distances are lost with coarser subsampling.
The distance errors are reflected in ATE$_{pos}$ because ATE calculates the RMSE over the whole path; there is no global reference in LOAM, and the early small errors can contribute to ATE more than the later errors~\cite{zhang2018tutorial, kummerle2009measuring, burgard2009comparison}.
For RE$_{pos}$, it calculates the averaged errors of separate sub-trajectories, and the distance errors are not accumulated over the whole path.

\noindent{\bf Summary.\quad}
Based on the experiments, we demonstrate that FLiCR is suitable for enabling edge-assisted online perceptions to mobile users.
Compared to the existing LiDAR point cloud compressions, it is {\em fast} in terms of the end-to-end compression/decompression latency, and {\em lightweight and efficient} in terms of energy usage and compression ratio.
%Moreover, it has advantages with low latency for the latency-performance tradeoff of online perceptions.
FLiCR achieves these benefits by affecting the quality of the point clouds using RI quantization and subsampling errors, and the end-to-end experiments of 3D object detection and LOAM show the impacts of the quality degradation %are dependent
on the downstream perception algorithms and their parameters.
Even though the lossy RIs have a different effect on the perception performance based on  each algorithm setting, ePSNR is able to quantify the point-wise and entropy-wise quality of a point cloud effectively.
Thus, when optimizing the compression method, it would be crucial to co-design the compression system with  awareness of the impact on downstream perceptions, and we leave this for future work.

%------------------------------------------------------------------------------
\section{Related Work}
\label{sec:related}
%------------------------------------------------------------------------------

Given the popularity of 3D point clouds, there are many point cloud compression methods.
Firstly, there are MPEG standard specifications: video-based point cloud compression (V-PCC) and geometry-based PCC (G-PCC)~\cite{graziosi2020overview}.
V-PCC converts 3D point clouds into 2D frames and compresses the frames with MPEG video codecs.
G-PCC directly leverages the octree structure as the intermediate representation (IR), and compresses the octree of point clouds.
Other than G-PCC, Google Draco~\cite{draco} and Point Cloud Library (PCL) compressors~\cite{rusu20113d} utilize tree structures including k-d tree and octree.
After generating the tree structure from a point cloud, the occupancy information with the leaf nodes is coded, and entropy or arithmetic coding is applied to compress the coded information~\cite{schnabel2006octree, devillers2000geometric}.
For range image compression, Tu \emph{et al.} present  direct mapping of sensor data to 2D frames by each laser ID with precision and compress these raw RIs using image compression methods~\cite{tu2016compressing}.
Other RI-based compression methods convert the raw sensor data from Cartesian coordinates into spherical coordinates by using the LiDAR sensor design~\cite{ahn2014large, feng2020real, houshiar20153d}.
Feng \emph{et al.} propose spatial encoding in the plane granularity and temporal optimization with scene alignment and prediction by using IMU fusion.
Even though these existing compression methods show decent compression performance, it is hard to apply them to our target use case of online remote perceptions because of their high latency magnitudes, as described in Section~\ref{sec:motivation}.

Recently, there has been research to utilize machine learning (ML) for LiDAR point cloud compression.
One popular approach is with the octree because the high compression ratio can be achieved by coding the tree into a more compact bytestream with  well-predicted occupancy information of a given tree~\cite{schnabel2006octree}.
By fully utilizing the relationship of neighboring nodes in the octree, the state-of-the-art works train the ML models to predict the distribution of the octree nodes~\cite{que2021voxelcontext, nguyen2021multiscale, biswas2020muscle, huang2020octsqueeze}.
With the predicted distribution, the occupancy information and nodes are effectively coded by assigning proper bits to each node of non-empty child nodes.
For RI-based ML approaches, the spatial optimization is done by using the encoder and decoder networks trained with RIs of point clouds~\cite{tu2019point, tu2019real}.
Some of these ML algorithms achieve sufficiently low latency to run in real-time~\cite{que2021voxelcontext, huang2020octsqueeze}. However, they are not practical for mobile users, because they rely on high-end processors and GPUs, which are usually unavailable for mobile devices.
Even if a mobile device has such computing resources, there is another issue with its limited battery.

%------------------------------------------------------------------------------
\section{Limitations and Future Work}
\label{sec:limitfuture}
%------------------------------------------------------------------------------

Although we show the effectiveness of FLiCR and ePSNR, there are still some remaining  limitations.
Firstly, as we observed with the end-to-end experiments, perception models pre-trained with the original data lose their predictive performance when used with point clouds reconstructed from lossy RIs.
To alleviate this issue, there is an opportunity to make the perception models robust to point clouds from different RI resolutions.
Another opportunity is to develop  dedicated hardware logic for the processing steps in Figure~\ref{fig:rioverview}.
As shown in Figure~\ref{fig:latbreak}, the RI conversion takes a large portion of the end-to-end latency.
Accelerating the conversion process would further improve the latency benefits of FLiCR. 
In addition, ePSNR has a limitation.
While ePSNR as a single-number metric effectively represents the point-wise and entropy-wise point cloud qualities, it requires two parameters: $\alpha$ and $\beta$.
We manually set these parameters for our experiments, but it is not scalable.
Therefore, there is a need to further develop a tuning methodology for these parameters, or to further refine the quality metric for LiDAR point clouds.
%------------------------------------------------------------------------------
\section{Conclusion}
\label{sec:conc}
%------------------------------------------------------------------------------
We describe the limitations of the existing point cloud compression methods for enabling LiDAR online perception on the edge.
We propose a lightweight, low-latency, and efficient compression method by using RI and dictionary coding.
For achieving the requirements, FLiCR fully leverages lossy RIs with quantization and subsampling.
To quantify the quality loss by quantization and subsampling, we introduce a new metric, ePSNR, which reflects both the point-wise and entropy-wise qualities of a point cloud.
We evaluate our compression method and demonstrate FLiCR is more appropriate for edge-assisted LiDAR online perceptions than the state-of-the-art compression algorithms.
Compared to the existing algorithm most suitable for the target use case, FLiCR takes up to 80 percent less end-to-end latency while presenting 12\mul compression ratio.
Our evaluation results with 3D object detection and LOAM show the impact of lossy RIs on the downstream perceptions and the effectiveness of ePSNR compared to the current quality metrics to capture this impact.

%------------------------------------------------------------------------------
\section*{Acknowledgment}
\label{sec:ack}
%------------------------------------------------------------------------------
We would like to thank the anonymous reviewers.
We are grateful to José Araújo, Héctor Caltenco, Bob Forsman, Per-Erik Brodin, and Gregoire Phillips for providing valuable feedback on this work and helping us improve its presentation.
This work has been partially supported by NSF projects CCF-2217070 and CNS-1909769, the Applications Driving Architectures (ADA) Research
Center, a JUMP Center co-sponsored by SRC and DARPA, and by funding and equipment gifts from VMware and Intel.

\bibliographystyle{IEEEtran}
\bibliography{sample-base}

\end{document}